\documentclass[12pt,preprint]{aastex}

\shorttitle{The First Extrasolar Planet with New Doppler Instrument}
\shortauthors{Ge et al.}

\begin{document}


\title{The First Extrasolar Planet Discovered with a New Generation
High Throughput Doppler Instrument}


\author{Jian Ge\altaffilmark{1}, Julian van Eyken\altaffilmark{1,2},
Suvrath Mahadevan\altaffilmark{1,2}, Curtis DeWitt\altaffilmark{1}, Stephen
R. Kane\altaffilmark{1}, Roger Cohen\altaffilmark{1},
Andrew Vanden Heuvel\altaffilmark{1}, Scott W. Fleming\altaffilmark{1} \&
Pengcheng Guo\altaffilmark{1}}
\affil{Department of Astronomy, The University of Florida, Gainesville, FL
32611}

\author{Gregory W. Henry}
\affil{Center of Excellence in Information System, Tennessee State
University, 3500 John A. Merritt Blvd., Box 9501, Nashville, TN 37209}

\author{Donald P. Schneider \& Lawrence W. Ramsey}
\affil{Department of Astronomy \& Astrophysics, The Pennsylvania State
University, University Park, PA 16802}

\author{Robert A. Wittenmyer, Michael Endl \& William D. Cochran}
\affil{Department of Astronomy, The University of Texas, Austin, TX 78712}

\author{Eric B. Ford}
\affil{Department of Astronomy, The University of California, Berkeley, CA
94720}

\author{Eduardo L. Mart\'in, \& Garik Israelian}
\affil{Instituto de Astrofisica de Canarias, La Laguna (Tenerife), Spain}

\author{Jeff Valenti}
\affil{Space Telescope Science Institute, 3700 San Martin Drive, Baltimore,
MD 21218}

\and 

\author{David Montes}
\affil{Departamento de Astrof\'isica y Ciencias de la Atm\'osfera, Facultad de F\'isicas, Universidad Complutense de Madrid, E-28040 Madrid, Spain}

\email{jge@astro.ufl.edu}

\altaffiltext{1}{Visiting Astronomer, Kitt Peak National Observatory.
KPNO is operated by AURA, Inc.\ under contract to the National Science
Foundation.}
\altaffiltext{2}{Michelson Graduate Fellow}


\begin{abstract}

We report the detection of the first extrasolar planet, ET-1 (HD 102195b),
using the Exoplanet Tracker (ET), a new generation Doppler instrument. The
planet orbits HD~102195, a young star with solar metallicity that may be 
part of the local association.  The planet imparts radial velocity 
variability to the star with a semiamplitude of $63.4\pm2.0$~m~s$^{-1}$ 
and a period of 4.11 days.  The planetary minimum mass ($m \sin i$) is 
$0.488\pm0.015$~$M_J$.  

The planet was initially detected in the spring of 2005 with the Kitt Peak 
National Observatory (KPNO) 0.9~m coud\'e feed telescope. The detection
was confirmed by radial velocity observations with the ET at the KPNO 2.1~m 
telescope and also at the 9~m Hobby-Eberly Telescope (HET) with its High 
Resolution Spectrograph.  This planetary discovery with a 0.9~m telescope 
around a $V=8.05$ magnitude star was made possible by the high throughput 
of the instrument: 49\% measured from the fiber output to the detector. The 
ET's interferometer-based approach is an effective method for planet 
detection. In addition, the ET concept is adaptable to multiple-object 
Doppler observations or very high precision observations with a 
cross-dispersed echelle spectrograph to separate stellar fringes over a 
broad wavelength band.
 
In addition to spectroscopic observations of HD~102195, we obtained 
brightness measurements with one of the automated photometric telescopes 
(APTs) at Fairborn Observatory. Those observations reveal that HD~102195 is 
a spotted variable star with an amplitude of $\sim0.015$~mag and a
$12.3 \pm 0.3$ day period. This is consistent with spectroscopically 
observed Ca~{\sc  ii} H~and~K emission levels and line broadening measurements 
but inconsistent with rotational modulation of surface activity as the 
cause of the radial velocity variability.  Our photometric observations 
rule out transits of the planetary companion.

\end{abstract}





\section{Introduction}

Over the past fifteen years, the field of extrasolar planets has moved 
from the fringes of science to become a central pillar of current and 
future astronomical studies. Although the first extrasolar planets were 
discovered by radio observations of a pulsar (Wolszczan \& Frail~1992), 
the vast majority of the over 170 known extrasolar planets orbit main 
sequence stars and were found using cross-dispersed echelle spectrographs 
at a dozen ground-based telescopes. The first detected extrasolar planet 
associated with a main sequence star, 51~Peg (Mayor \& Queloz 1995), 
ushered in a continuous stream of unexpected results on extrasolar planets, 
ranging from their extreme diversity (``hot Jupiters", planets in very 
elongated orbits, multiple-Jupiter-mass planetary systems) to the recently 
discovered super-Earth-mass planets around solar-type stars with orbital
periods of a few days. A review of the field is given by Marcy et al. (2005). 
These discoveries not only provide new challenges for the fields of planetary 
origins and evolution, but also indicate that a large sample of planets is 
required to obtain a full understanding of their nature.

Although the high-precision echelle Doppler instruments have proven quite 
successful at detecting extrasolar planets, the current approach is costly 
and time-consuming because of the large telescopes required and the 
relatively low throughputs (a few percent) of the spectrographs as well 
as the limitation of observing one star at a time. A sample of approximately 
5,000 stars (generally the closest and brightest ones), including the 2000 N2K survey stars targeting 
for short period planets (Fischer et al. 2005), has been monitored 
for planets with echelle instruments on a dozen telescopes, including most 
of the new generation large telescopes such as the Keck, the Very Large 
Telescope (VLT), Subaru, the Hobby-Eberly Telescope (HET), and Magellan.  
Most of the target stars have visual magnitudes brighter than about 8.0 except the 
N2K targets stars with visual magnitudes brighter than about 10.5 (Fischer et al. 2005).  
Based on the current planetary detection rate of $\sim~7\%$ among solar type 
stars (Marcy et al. 2005), a few hundred planets will likely be detected 
over the next ten years using current techniques. Given the surprising 
range of properties known to date (and the likelihood of even larger 
surprises in the future), this sample will still be inadequate for giving us 
a comprehensive picture of planet formation and evolution and their relation 
to stellar properties such as mass, luminosity, spectral type, metallicity, 
duplicity, and age. Furthermore, to monitor a large number of stars, one 
must move to targets with $V > 8.0$.  An instrument with high throughput 
becomes critical to search for planets around fainter stars.  In addition, 
an instrument with multiple object capability would greatly enhance the 
efficiency of any extrasolar planet survey.

A promising Doppler technique for finding extrasolar planets, one that is 
quite different from the echelle approach, uses a dispersed fixed-delay 
interferometer (DFDI) for precision radial velocity measurements. The DFDI 
offers high throughput and multi-object capability.  Instead of measuring 
the absorption line centroid shifts in the echelle approach, a DFDI determines 
the radial velocity by monitoring interference fringe phase shifts.  The 
idea for using a fixed-delay interferometer for high precision Doppler 
measurements was first proposed by solar astrophysicists in the 1970s and 
1980s to measure solar oscillations (Barker \& Hollenbach 1972; Gorskii \& 
Lebedev 1977; Beckers \& Brown 1978, Kozhevatov 1983). This approach was 
adopted by the Global Oscillations Network Group (GONG) interferometer 
(Harvey et al. 1995).  The GONG interferometer, with a narrow passband 
($\sim~1$ \AA), has produced very high Doppler precision measurements of the 
sun (sub m~s$^{-1}$ precision for the GONG measurements; Harvey~2002, private 
communication).  The concept of combining a fixed-delay interferometer with 
a moderate resolution spectrometer for broad band operations for high 
precision stellar Doppler measurements was proposed in 1997 by David Erskine 
at Lawrence Livermore National Lab.  The initial lab experiments and telescope 
observations successfully demonstrated the DFDI concept [earlier this concept 
was called fringing spectrometer and Externally Dispersed Interferometer 
(EDI), Erskine \& Ge 2000; Ge, Erskine \& Rushford 2002]. The theory for 
DFDI is described in Ge~(2002).  In 2002 a prototype DFDI instrument, 
designated as the Exoplanet Tracker (ET), was used at the KPNO 2.1~m 
telescope to reproduce the radial velocity curve of the known planet 
orbiting the solar type star 51~Peg (van Eyken et al. 2004a), demonstrating 
the capability of this new approach for planet detection.  A new version of 
the ET instrument, optimized for high throughput and relatively large 
wavelength coverage compared to the 2002 prototype, was commissioned at 
the KPNO 0.9~m coud\'e feed and 2.1~m telescopes in 2003 November (Ge et al. 
2004; van Eyken et al. 2004b).  The first DFDI multi-object observations, 
using a modified ET, were obtained in 2005 March at the 2.5~m Sloan Digital 
Sky Survey (SDSS) telescope (York et al. 2000; Gunn et al. 2006) at Apache 
Point Observatory (Ge et al. 2005).

In this paper, we report the first detection of a new extrasolar planet 
using the DFDI technique. The planet, HD~102195b (ET-1), associated with 
the star HD~102195, has an orbital period of 4.11~days and was discovered 
with ET at the KPNO 0.9~m coud\'e feed and 2.1~m telescopes.  The planet 
survey instrumentation and observations are reviewed in section~2, and a 
description of the survey data processing is provided in section~3.  
Section~4 presents additional radial velocity, spectroscopic, and 
photometric observations of HD~102195. An analysis of the radial 
velocity curve  and follow-up data of HD~102195
 and a brief discussion of the properties of 
ET-1 are given in sections~5 and~6, respectively.

\section{Exoplanet Tracker:  Description and Initial Survey}


In the winter of 2004 we began a small-scale extrasolar planet survey
at KPNO using the DFDI approach. This program is designed to detect
new planets with short orbital periods \hbox{($<$ 10 days).} Targets
were chosen primarily from the Nstar catalogue (Gray et al. 2003),
selecting dwarf stars of type FGK with \hbox{$7.8 < V < 9$.} Stars
that were known to be fast rotators or had high activity indicators were
removed from the sample, as were any known visual doubles or
variables. Since studies have shown a strong correlation between
frequency of planetary systems and high metallicity in the star (e.g.,
Gonzalez 1997; Reid 2002; Santos et al. 2004; Fischer \& Valenti
2005), we selected relatively high metallicity stars \hbox{([M/H]$>$0.0.)} 
to increase the planet detection efficiency.

Below we present a brief outline of the DFDI instrument, ET, of which a more detailed description 
can be found in van Eyken et al~(2004b). 
 ET includes a single object fiber feed 
system, a Michelson type interferometer with a fixed optical delay in one of the arms,
 a spectrograph with a Volume Phase Holographic 
(VPH) grating, and a 4k$\times$4k CCD camera with 15~$\mu$m pixels.  An 
optical fiber with a 200~$\mu$m core diameter (2.5$''$ on the sky) is fed 
by an $f/8$ beam from the KPNO 0.9~m coud\'e feed or 2.1~m telescope (both 
telescopes are housed in the same enclosure and feed the same spectrograph, 
so the instrument does not have to be relocated). The fiber has an $f/6$ 
output beam.  ET is designed to operate from 5000--5640 \AA\ and has a 
spectral resolving power of $\sim$5,100. A resolution element is sampled 
by 6.7~pixels in the dispersion direction. Each fringe is sampled by
$\sim58$ pixels in the slit direction; this range usually includes 
$\sim5$ periods of fringing.  The VPH grating is a Dickson type design, which 
 produces a 92\% peak grating efficiency but with a limited grating 
operation band of $\sim600$ \AA. The CCD camera was purchased from Spectral 
Instrument Inc. The detector is a back-illuminated CCD with $\sim90$\% 
quantum efficiency in the ET operating wavelengths.

The measured instrument throughput from the fiber output to the detector 
is 49\%. The overall average detection efficiency, including the telescope, 
seeing, fiber, instrument and detection losses, is 18\% under typical seeing 
conditions ($\sim$~1.5$''$) at the KPNO coud\'e feed/2.1~m.  An iodine vapor 
glass cell 150~mm long and 50~mm in diameter is used as a Doppler 
zero-velocity reference. The cell temperature is stabilized to 
\hbox{$60 \pm 0.1^\circ$C.} Since 2003, we have been able to recover the 
expected RV signatures of known planets routinely with the ET instrument
(see Figure~1) and have obtained short term (two day) precision as high as 
3.6~m~s$^{-1}$ (photon noise limited) on the bright RV stable star
36~UMa (van Eyken et al. 2004b). The instrument long term precision has not been 
well chacterized yet. The preliminary RV measurements over two months show that the 
instrument is at least stable to $\sim$ 13~m~s$^{-1}$. 

The initial ET planet survey of 90 stars without previous precision radial
velocity observations was conducted during the period 2004 December--2005 
May with the KPNO 0.9~m coud\'e feed from December to March and the 2.1-m 
telescope in May. A total of 59 nights of the coud\'e feed time and 7 nights 
of 2.1-m time were allocated to the program.  Observations on 43 nights were 
obtained during the survey (losses were primarily due to poor weather).

The survey was divided into five observation blocks, each of $\sim$~10 
nights duration.  Star and iodine templates were obtained at the beginning 
and end of each block (van Eyken et al. 2004a). For a typical block, 5--6 
RV measurements were acquired for each survey star.  The RV data were taken 
with the iodine vapor cell in the stellar beam. The instrument was set to 
the same configuration during each observation block to maximize the 
instrument stability. The interferometer fringe pattern was stablized by 
a closed-loop, actively-controlled piezoelectric transducer (PZT) system 
connected to one of the interferometer arms. The fringe pattern was 
monitored with a stablized HeNe laser at 0.6328~$\mu$m purchased from Melles 
Griot Inc. To measure the illumination profile and the spectral line slant 
for each fringe while maximizing the instrument stability, the interferometer 
fringes were jittered at the beginning and the end of each observation 
block to produce non-fringe calibration data. The jitter of the interferometer 
fringes was produced by operating the interferometer PZT ramp generators at 
about 10~Hz. The interferometer phase was locked during the observations 
in each block. The instrument room was heated to $\sim$24$^\circ$C, and the 
rms temperature fluctuation during the runs was approximately 0.1$^\circ$C. 
The absolute velocity drift due to the temperature fluctuation and 
mechanical instability is about 1--2 km~s$^{-1}$. This velocity drift 
is calibrated and corrected with the iodine absorption line fringes.

Typical individual exposure times are $\sim$25 min for $V\sim8$ and $\sim$40 
min for $V\sim9$ stars at the coud\'e feed; typical exposure times for the 
2.1~m are 10~min for all targets.  The average rms RV precision (photon 
noise) \hbox{is $\sim20$ m s$^{-1}$} at the coud\'e feed \hbox{and 
$\sim17$ m s$^{-1}$ at the 2.1~m.} A total of $\sim$650 ET observations of 
90 stars were obtained during this campaign.




\section{Survey Data Processing}

The data produced by ET required the development of a large software 
processing system. In this section we briefly outline the steps required to 
extract precision RV measurements from the raw data recorded by the detector.  
The main steps in the processing of the ET fringe spectra are (1)~image 
pre-processing, (2)~sinusoid fitting, and (3)~RV shift determination.

\noindent{\bf Preprocessing}: Image preprocessing is performed with a
combination of standard IRAF procedures\footnote{IRAF is distributed by the National Optical Observatory,
which is operated by the Association of Universities for Research in
Astronomy, Inc., under contract with the National Science Foundation.}
 and proprietary software written in
Research Systems Inc's IDL data analysis language. The steps consist of 
the following:

\noindent Background subtraction/trimming:
Detector bias and background light calibration images (dark plus stray light)
 are first subtracted 
from the raw data.  An attempt is made to remove as much internally 
scattered light as possible by fitting a smooth function to the dark areas 
of the images and using a one-dimensional interpolation scheme to estimate 
the scattered light levels in the areas covered by the spectra.  The 
two-dimensional frames are then trimmed into new two-dimensional frames 
including only fringe spectra.  

\noindent Flatfielding: Pixel-to-pixel sensitivity variations are corrected by
dividing the data by a flatfield calibration image taken with a tungsten
lamp. The irregular illumination of the spectrum is repaired by the 
application of a `self illumination correction' algorithm. This process 
constructs a second flatfield by extracting the underlying illumination 
function from each individual image (see van Eyken et al.~2004a).

\noindent Slant correction: An algorithm is applied to correct for
misalignment of the spectral lines with the rows/columns of the CCD.
This feature is caused by a combination of imperfect alignment of the 
CCD and by aberration and distortion in the instrument optics (i.e., a 
simple rotation will not correct the alignment over the entire detector). 
This step is essential to obtain the proper cuts along the wavelength 
channels required to obtain uniform fringes.

\noindent Low pass filtering: Finally, a one-dimensional, low-pass Fourier
filter is applied to the data in the dispersion direction. This action
removes the interferometer comb, the pattern of parallel lines that is 
created by the continuum and therefore contains no Doppler information.

\noindent {\bf Phase and Visibility Determination}: Once the initial image
processing is complete, the phases and visibilities of each wavelength 
channel (column) are measured. This is achieved simply by fitting a sine wave to 
each wavelength channel in the image with a standard $\chi^2$ minimization 
algorithm. The fit is weighted according to the number of counts in the 
original non-flatfielded data, on the assumption of photon noise 
dominated error.

To determine the phase accurately, two passes of the curve-fitting
are performed to determine the fringe frequency. The first pass
determines the frequencies of each channel. A polynomial is then fit to
the frequencies as a function of wavelength (weighted according to the 
measurement errors). The process is repeated a second time but with the
frequencies fixed to match this function. This approach was found to
significantly improve the final precision.

\noindent {\bf Determining the Intrinsic Doppler Shift}: Once the phase and 
visibility values are obtained for all wavelength channels, we combine them 
to form a vector vs. wavelength channel called a whirl, where each vector 
represents the fringe amplitude and phase (see the definition in Erskine \& 
Ge 2000).  In general, the star+iodine data whirl will be a linear vector 
combination of the iodine and star template whirls, where the template whirls 
are rotated in phase by an amount that must be determined. A set of four 
simultaneous linear equations can be constructed for each pair of channels (Erskine 
2003). 
The solution for the phase rotations is found by linearizing the complete 
overdetermined set of equations across all channels and using singular 
value decomposition. This procedure returns a $``$best fit solution" along 
with standard error estimates.  The measured rotations correspond to the 
star shift and the intrinsic instrument shift. The difference between these 
two rotations yield the intrinsic stellar RV shift.

In addition to the two phase rotations, several additional degrees of freedom
are allowed in our processing.  In particular, we allow for bulk shift of 
the entire spectrum in the dispersion direction.  This change can be 
produced by a Doppler shift or by movement of the CCD detector due to thermal 
flexure. This, along with additional flexures in the instrument optics, 
can create translations of the image.  A reduced $\chi^2$ value is determined 
between the star+iodine data and the best superposition of the pixel-shifted 
and rotated templates.  This process is iterated until a minimum is reached 
in the reduced $\chi^2$.

After correcting for instrument drift, it still remains to apply a barycentric 
correction to the velocities to account for the motion of the Earth. This is done using our
own software written in IDL. Diurnal motion
is corrected using an algorithm adapted from the IRAF procedure 'rvcorrect' in the NOAO
package. Annual motion is corrected using the 'baryvel' routine from the 
IDL astronomy library (ref. http://idlastro.gsfc.nasa.gov), which is based on
the algorithm from Stumpff (1980). 

At the completion of the data processing, the 90 stars in the survey fell 
into three catagories: (1) 75 of the stars exhibited less than 2.5$\sigma$ 
RV scatter about a mean value; (2) 10 stars had RV variations between 
2.5$\sigma$ and 1000~m~s$^{-1}$ rms about the mean; and (3) 5 stars were 
spectroscopic binaries with RV variations larger than 1000 m s$^{-1}$ rms. 
Only stars in the second group were considered candidates for harboring a
planet and, of those ten stars, HD~102195 appeared to be the most promising 
candidate.

\section{HD~102195}

Having identified HD~102195 as the most promising planetary-candidate star
from our initial Exoplanet Tracker survey, we embarked on a series of 
additional observations of HD~102195 to determine the nature of the RV 
variations.  We obtained further precision RV measurements as well as 
high-resolution spectroscopy, Ca~{\sc  ii} H~and~K emission-line spectroscopy, 
and high-precision photometry.

\subsection{Additional Radial Velocities}

The initial 14 coud\'e measurements of HD~102195 with the ET had an rms 
variation of slightly more than 60~m~s$^{-1}$, with typical single-measurement 
errors of 20~m~s$^{-1}$.  An additional 14 ET measurements were made in 
2005 May with the KPNO 2.1~m telescope and the same ET instrument.  
Figure~2 displays all 28 spring 2005 ET radial velocities of the star and 
confirms HD~102195 as a good planetary candidate host star.

The KPNO 2.1~m ET system was used again in 2005 December to obtain an 
additional 21 RV measurements of HD~102195. We also acquired 10 RV 
measurements of the star between 2005 November and 2006 January with 
the High Resolution Spectrograph (HRS) of the Hobby-Eberly Telescope (HET)
(Ramsey et al. 2006) to confirm the ET results.  The HET/HRS data were 
obtained with a 2$''$ fiber and exposure times of 10 minutes. These
observations were conducted with the HRS $R = 60,000$ cross-dispersed
echelle mode. This is the standard HRS configuration for precision
RV measurements, and the data were processed with the techniques
described in Cochran et al. (2004).  The HET observations, which have typical 
internal errors of \hbox{$\sim$2 m s$^{-1}$,} were able to be timed for 
maximum phase leverage because of the queue-scheduled operation of the 
telescope. All 59 precision RV measurements of HD~102195 from all sources
are presented in Table~1.

\subsection{High-Resolution Spectra}

A total of nine high-resolution ($R=150,000$) spectra of HD~102195 were 
obtained with the SARG spectrograph on the 3.5~m TNG telescope at La Palma
on 2005 June 19, 20, and 21. The wavelength coverage is complete from 3700 \AA\ 
to 10000 \AA. These data were used to monitor line-bisector 
variations, to determine stellar properties (metallicity, $\log g$, 
T$_{\rm eff}$ and $v \sin i$), and to search for any evidence of a second 
set of lines in the system.  The SARG spectra were processed through a 
data pipeline developed by J.~Valenti for the N2K Consortium short-period 
planet survey (Valenti \& Fischer 2005).  Figure~3 shows a section
of the SARG data centered at a wavelength of~6140~\AA. The typical 
signal-to-noise ratio (S/N) for each spectrum is about 100 per pixel. The 
nine spectra were normalized and fitted with synthetic spectra to derive 
the stellar parameters. The average values of the derived parameters are 
reported in section 5.1. The nine spectra were searched for line bisector 
variations, but no significant variations were found over the three days. For instance, the average bisector velocity span for each day is 
 $-1.9\pm12.4$ ms$^{-1}$, $5.2\pm12.0$ ms$^{-1}$, and $23.6\pm11.7$ ms$^{-1}$ for June 19, 20 and 21, respectively. Details on the line bisector analysis can be seen in Martinez Fiorenzano et al. 
(2005). ,

In addition, one high resolution spectrum of HD 102195 was obtained with
the 2.2~m telescope at the German Spanish Astronomical Observatory (CAHA)
(Almer\'{\i}a, Spain) on 2006 January 14. The Fibre Optics Cassegrain
Echelle Spectrograph (FOCES) (Pfeiffer et al. 1998) was used with a
$2048\times2048$ 24$\mu$m SITE$\#$1d15 CCD detector. The wavelength range covers
from 3500 to 10700 \AA\ in 111 orders. The reciprocal dispersion ranges
from 0.04 to 0.13 \AA/pixel and the spectral resolution, determined as the
full width at half maximum (FWHM) of the arc comparison lines, ranges from
0.08 to 0.35 \AA. A signal to noise ratio S/N=100 per pixel was obtained in the
H$\alpha$ line region.

The spectra have been extracted using the standard
reduction procedures in the
IRAF package (bias subtraction,
flat-field division and optimal extraction of the spectra).
The wavelength calibration was obtained by taking
spectra of a Th-Ar lamp.
Finally, the spectra have been normalized by
a polynomial fit to the observed continuum.

This spectrum was used to determine stellar properties, analyze the
chromospheric activity, and estimate the age from the Lithium line (Li~{\sc i}
6707.8).

\subsection{Ca II H and K Observations}

HD~102195 was observed between 2005 December 14--20 with the KPNO 0.9~m 
coud\'e spectrograph to monitor the emission at the core of the Ca~H and~K 
lines.  The observations were obtained with Grating~A, Camera~5, the long 
collimator, a 400~$\mu$m slit, and a 2k$\times$2k CCD camera.  This 
configuration produced a spectral resolving power of 10,000 over a 
wavelength range of 230~\AA\ centered on 4010~\AA.  A total of 29 spectra 
were obtained during the run.  The spectra were processed with standard 
IRAF procedures.  A reference star, $\tau$~Ceti, was monitored for 
comparison.  Emission is clearly seen in the core of the Ca~{\sc  ii} H~and~K 
lines (see Figure~4), as expected by the classification of this star as 
mildly active by Strassmeier et al. (2000).

\subsection{High-Precision Photometry}

Between 2005 April and 2006 February, we obtained high-precision photometry 
of HD~102195 with the T10 0.8~m automatic photometric telescope (APT) at 
Fairborn Observatory.  These observations covered the last part of the 
2004--05 and the first part of the 2005--06 observing seasons.  The APTs 
can detect short-term, low-amplitude brightness variability in the stars 
caused by rotational modulation in the visibility of magnetic surface 
features such as spots and plages (e.g., Henry, Fekel, \& Hall 1995) as 
well as longer-term variations associated with stellar magnetic cycles 
(Henry 1999).  Thus, photometric observations can help to establish whether 
observed radial velocity variations are caused by stellar activity or 
planetary-reflex motion (e.g., Henry et al. (2000a).  Queloz et al. (2001)
and Paulson et al. (2004) have found several examples of periodic radial 
velocity variations in solar-type stars caused by photospheric spots and 
plages.  The APT observations are also useful to search for possible 
transits of the planetary companions (Henry et al. 2000b; Sato et al. 2005).

The T10 APT is equipped with a two-channel precision photometer containing 
two EMI 9124QB bi-alkali photomultiplier tubes to make simultaneous 
measurements in the Str\"omgren $b$ and $y$ passbands.  The APT measures 
the difference in brightness between a program star and one or more nearby 
comparison stars.  The primary comparison star used for HD~102195 was 
HD~102747 ($V$ = 7.77, $B-V$ = 0.513, F5); a secondary comparison star was 
HD~101730 ($V$ = 6.94, $B-V$ = 0.49, F5).  Str\"omgren $b$ and $y$ 
differential magnitudes were computed and corrected for differential 
extinction with nightly extinction coefficients and transformed to the 
Str\"omgren system with yearly mean transformation coefficients.  Finally, we 
combined the Str\"omgren $b$ and $y$ differential magnitudes into a single 
$(b+y)/2$ passband to maximize the precision of the photometric measurements.  
The typical external precision of the differential magnitudes is 
0.0012--0.0016 mag for this telescope, as determined from observations of 
pairs of constant stars.  The standard deviation of our comparison star 2 
minus comparison star 1 differential magnitudes is 0.0019, close to the 
typical precision and indicating very little variability in either comparison 
star.  However, the HD~102195 minus comparison star 1 differential magnitudes 
have standard deviations of 0.0048 and 0.0033 mag in observing seasons 1 and 
2, respectively, indicating clear variability in HD~102195.  The 468 
individual differential magnitudes of HD~102195 minus the primary comparison 
star are given in Table 2.  Further information on the automatic telescope, 
photometer, observing procedures, and data reduction techniques can be found 
in Henry (1999) and Eaton, Henry, \& Fekel (2003).

\section{Results} 

\subsection{Stellar Properties}

HD~102195 was identified as a K0V star with a color of $(B-V)$ = +0.84 
(Strassmeier et al. 2000).  The {\it Hipparcos} parallax (ESA 1997) of 
34.51$\pm$1.16 mas places this target at 29~pc.  The apparent visible 
magnitude is $V = 8.05\pm0.03$ measured from the APT photometry.  The 
absolute magnitude is $M_V=5.73$. Montes et al. (2001) suggest that 
HD~102195 may belong to the Local Association, which, if confirmed, would place 
the age of the star at 20--150~Myr. Following the procedure described 
in Valenti \& Fischer~(2005), our spectroscopic analysis of the SARG 
spectra ($R \sim 150,000$) yields \hbox{$T_{\rm eff} = 5330 \pm 28$K,} 
\hbox{[Fe/H]= 0.096$ \pm 0.032$,} \hbox{$\log g$ = 4.368 $\pm 0.038$ 
[log(cm s$^{-2}$)],} and \hbox{$v \sin i= 3.23 \pm 0.07$ km s$^{-1}$.}  The 
fit of this model to the combined spectrum is shown in Figure~3. The \
effective temperature and surface gravity of the star appears to be 
consistent with a G8V.

We interpolated the $``$high temperature" table of VandenBerg \&
 Clem (2003) as a function of spectroscopic effective temperature,
 gravity, and iron abundance to obtain a V-band bolometric correction
 of $-0.177$. Applying this correction to the observed V-band
 magnitude of $8.05 \pm 0.03$ yielded a stellar luminosity of
 $0.463 \pm 0.034$ L$_\odot$.

 We used our spectroscopically determined [Ti/Fe] abundance ratio
 as a crude measure of $\alpha$-element enrichment, obtaining
 $0.049 \pm 0.042$. We detect no significant $\alpha$-element
 enhancement, but for consistency with Valenti \& Fischer (2005),
 we used our measured $\alpha$-element enhancement of 0.049 when
 interpolating the Y$^2$ isochrones, slightly perturbing our
 derived stellar properties.

 We used the stellar luminosity and the spectroscopic effective
 temperature, iron abundance, and $\alpha$-element enrichment to
 interpolate the Y$^2$ isochrones (Demarque et al. 2004),
 obtaining a stellar mass of 0.926 $M_\odot$ and a stellar radius of 
0.835~R$_\odot$. In addition, 
the most probable age is 2.0 Gyr with an asymmetric 1$\sigma$ confidence 
interval spanning the range 0.6 to 4.2 Gyr. This age estimate is much older 
than that derived from the Local Association by Montes et al. (2001). The 
chromospheric activity of HD 102195 has been measured by Strassmeier et al. 
(2000). The chromospheric emission ratio, log $R'_{HK} = -4.30$, indicates 
this is a mildly active star (Noyes et al. 1984).  The stellar parameters 
are summarized in Table~3.

Further analysis of the FOCES high resolution optical spectrum of HD 102195 was 
conducted to independently derive the stellar properties. Here we summarize the results. 

\noindent{\bf Spectral Type}

To obtain an independent estimate of the spectral type of this star, we
have compared the spectrum of HD~102195 with that of inactive reference
stars taken during the same observing run. The analysis makes use of the
program {\sc starmod} developed at Penn State University (Barden 1985) and
modified more recently by us. With this program a synthetic stellar
spectrum is constructed from the artificially rotationally broadened, and
radial-velocity shifted spectrum of an appropriate reference star. We
obtained the best fit between observed and synthetic spectra when we use a
G8-dwarf spectral type standard star (HD~182488). The uncertainty in this
classification is of one spectral subtype as is typical in the MK spectral
classification. The careful analysis of the wings of the H$\alpha$ and
Na~{\sc i} D$_{\rm 1}$ and D$_{\rm 2}$ lines clearly indicates a better
fit with a G8V than with a K0V. This spectral classification agrees with
the results of the spectral synthesis of the SARG spectra.

\noindent {\bf Rotational velocity}

By using the program {\sc starmod} we have obtained the best fits with
$v\sin{i}$ values between 3 and 4 km s$^{-1}$. In order to determine a
more accurate rotational velocity of this star we have made use of the
cross-correlation technique by using the routine {\sc fxcor} in IRAF. When
a stellar spectrum with rotationally broadened lines is cross-correlated
against a narrow-lined spectrum, the width of the cross-correlation
function (CCF) is sensitive to the amount of rotational broadening of the
first spectrum. Thus, by measuring this width, one can obtain a
measurement of the rotational velocity of the star. The observed spectrum
was $cross-correlated$ against the spectrum of a template star (the G8V
star HD~182488) and the width (FWHM) of CCF determined. The rotational 
velocity of HD 182488 is 0.6 $\pm$ 0.5 km/s, this values has
been determined by Fekel (1997) by a method based in the
determination of the intrinsic stellar broadening from the observed FWHM
of weak or moderate-strength lines, corrected from instrumental profile
and from macroturbulent broadening.
The calibration of
this width to yield an estimation of $v\sin{i}$ is determined by
cross-correlating artificially broadened spectra of the template star with
the original template star spectrum. The broadened spectra was created for
$v\sin{i}$ spanning the expected range of values by convolution with a
theoretical rotational profile (Gray 1992) using the program {\sc
starmod}. The resultant relationship between $v\sin{i}$ and FWHM of the
CCF was fitted with a fourth-order polynomial. The $v\sin{i}$ obtained in
this way for HD~102195 is 3.7$\pm$0.6 km~s$^{-1}$ which is in good
agreement with the value derived from the SARG spectra.

\noindent {\bf Radial velocity}

The previous known value of the heliocentric radial velocity of this star
is 2.1$\pm$0.6 km~s$^{-1}$, a weighted mean value of the two measurements reported
by Strassmeier et al. (2000).

In the FOCES spectrum the heliocentric radial velocity has been determined
by using the cross-correlation technique. The spectrum of HD~102195 was
cross-correlated order by order, by using the routine {\sc fxcor} in IRAF,
against the spectrum of the K0V radial velocity standard HD~3651 (Barnes
et al. 1986). 
We have used the K0V star HD 3651 because the G8V star HD 182488 was not
observed in the same observing run as that for HD 102195. However, we have
determined also the radial velocity using HD 182488 and the result is very
similar. The radial velocity was derived for each order from the
 peak of the cross-correlation function peak (CCF), and the
uncertainties were calculated by {\sc fxcor} based on the fitted peak
height and the antisymmetric noise as described by Tonry \& Davis (1979).
Those orders which contain chromospheric features and prominent telluric
lines have been excluded when determining the mean velocity. The resulting
value is $V_{\rm hel}$ = 2.04$\pm$0.07 km~s$^{-1}$.

\noindent {\bf Kinematics and age}

The galactic space-velocity components ($U$, $V$, $W$) have been
determined by Montes et al. (2001) using heliocentric radial velocity of
2.1$\pm$0.6 km~s$^{-1}$ (Strassmeier et al. 2000) and the precise proper
motions taken from Hipparcos (ESA 1997) and Tycho-2 (H$\o$g et al. 2000)
catalogues. Solely on the basis of  kinematics criteria Montes et al. (2001)
classified HD~102195 as a young disk star and a possible member of the Local
Association moving group. The age of this complex moving group is in the
range of 20 to 150 Myr; however, the age inferred from the Lithium line strength does
not agree with such a  young age.

As it is well known, the Li~{\sc i} line at $\lambda$6708 \AA\ is an important
diagnostic of age in late-type stars, since it is destroyed easily by
thermonuclear reactions in the stellar interior (e.g. Soderblom et al. 1990). In the FOCES spectrum a small absorption Li~{\sc i} line is observed blended with the nearby
Fe~{\sc i} $\lambda$6707.41~\AA\ line. We have corrected the total
measured equivalent width, $EW$(Li~{\sc i}+Fe~{\sc i}) = 17.4 m\AA, by
subtracting the $EW$ of Fe~{\sc i} calculated from the empirical
relationship with ($B$--$V$) given by Soderblom et al. (1993). 
Very similar results are also obtained using the espiral relationship
 given by Favata et al. (1993). The
resulting corrected $EW$(Li~{\sc i}) is 2.8 m\AA, in agreement with the
value previously reported by Strassmeier et al. (2000) of 3 m\AA.

Comparing the $EW$(Li~{\sc i}) of HD~102195 with those of well-known young
open clusters of different ages in a  $EW$(Li~{\sc i}) versus spectral
type diagram (Montes et al. 2001 and references therein), a much greter age than the Local Association is inferred, it
could be even older than the Hyades. This result and the relatively low
level of chromospheric activity (see below) favor the range of age deduced
from the Y2 isochrones.

\noindent {\bf Chromospheric activity indicators}

The FOCES 01/14/06 echelle spectrum of HD~102195 allows us to study the
behaviour of the different chromospheric activity indicators from the
Ca~{\sc ii} H \& K to the Ca~{\sc ii} IRT lines, which are formed at
different atmospheric heights. The chromospheric contribution in these
features has been determined by using the spectral subtraction technique
(see Montes et al. 2000).  The spectral subtraction technique  is the subtraction of a
synthesized stellar spectrum constructed from artificially rotationally
broadened, and radial velocity shifted spectrum of an inactive star
chosen to match the spectral type and luminosity class of the active star
under consideration. The synthesized spectrum was constructed using a
G8V reference star with the program {\sc starmod}.

In the observed spectrum only the Ca~{\sc ii} H \& K lines are clearly
detected in emission. After applying the spectral subtraction a small
excess chromospheric emission is detected in the H$\alpha$ and Ca~{\sc ii}
IRT ($\lambda$8498, $\lambda$8542, $\lambda$8662) lines. However,
the other Balmer lines as well as the Na~{\sc i} D$_{\rm 1}$ and D$_{\rm
2}$ lines do not show evidences of filled-in absorption by chromospheric
emission.

In Table 4 we give the excess emission equivalent width
($EW$) (measured in the subtracted spectra) for the Ca~{\sc ii}~H~\&~K,
H$\alpha$ and Ca~{\sc ii} IRT lines and the corresponding absolute
chromospheric flux at the stellar surface (logF$_{\rm S}$ (erg cm$^{-2}$
s$^{-1}$)) obtained by using the calibration of Hall (1996) as a function
of ($B$--$V$). For instance, the calibration of Hall (1996) give the flux 
in the continuum at the wavelengths 3950 \AA\ (for the H\&K CaII lines), 
6563 \AA\ (H$_alpha$ line) and 8520 \AA\ (CaII IRT lines) as a function of B-V. 
Using this flux in the
continuum we convert the EW into logFs.
The uncertainties in the measured $EW$ were estimated to be
in the range 10-20\% taking into account:
a) the typical internal precisions of {\sc starmod},
b) the rms obtained in the fit between observed and synthesized spectra in
the spectral regions outside the chromospheric features
 and
c) the standard deviations resulting in the
$EW$ measurements.

The chromospheric activity index ($R'_{HK}$) obtained from the stellar
surface in the Ca~{\sc ii} H \& K lines (which is corrected from the
photospheric contribution by using the spectral subtraction) is
log$R'_{HK}$ = $-$4.45 similar to the value or $-$4.3 reported by Strassmeier
et al. (2000).

It is well known the chromospheric activity is related to both the
spectral type and rotational velocity. Late spectral type main sequence
stars have larger chromospheric emission than early stars and as the star
ages, it slows down its rotation and decrease the level of activity. In
this sense the chromospheric activity provides an indication of the
stellar age for a given spectral type. Using the calibration of Soderblom
et al. (1991) for the chromospheric activity (measured by $R'_{HK}$) --
age relation an age of 0.8 Gyr is inferred. This age is only an
estimation, since these kind of calibration are strictly valid for
chromospherically quiet stars (log$R'_{HK}$ $<$ -4.75) (see Saffe et al.
2005 for a discussion), but agrees with the age deduced from the Li~{\sc
i} line and it is in the range of age deduced from the Y2 isochrones.

\subsection{RV Analysis and Orbital Solution}

The RV precision and stability of ET have been determined using the RV 
stable stars in our survey.  We define RV stable stars (at our precision 
level) to be those stars that show a reduced $\chi^2 < 2$ to the fit of a 
constant RV value over the period of observation (typically $\sim$ 10 days). 
At the 0.9~m coud\'e during a typical observing block, 7 out of 15 search 
stars satisfied this criterion. For those, we found a rms scatter of 
18.9~m~s$^{-1}$ at a mean visual magnitude of 8.26 and typical exposure 
times of 20--30~min.  Using the same criterion for the 2.1~m observations, 
we found 12 out of 24 stars to be RV stable. These stable stars had a mean 
standard deviation of 17.6~m~s$^{-1}$ about a constant RV.  For the 2.1~m 
targets, which had $<V>$ = 8.48, the typical exposure times were 10 minutes.  
For both configurations, the errors are consistent with the expected photon 
noise limit. For an independent check of the instrument, we were able to 
recover the RV curve for 51~Peg. Figure~1 shows the coud\'e feed results 
for 51 Peg with the predicted RV curve due to the 4.23-day companion 
51~Peg b.  The measured residual mean error (after the expected velocity 
induced by the planet) for the coud\'e feed observations of 51~Peg is 
7.9 m s$^{-1}$, which is consistent with the photon noise limit.

A total of 59 radial velocities of HD~102195 have been obtained from 
2005 January through 2006 January.  In contrast with the stable stars, the 
velocity variability of HD~102195 is $\sim$ 60 m s$^{-1}$, 3$\sigma$ above 
the measurement noise for the ET measurements at the KPNO coud\'e, 6$\sigma$ 
above the noise level of the ET measurements at the KPNO 2.1m, and approximately 30 times 
larger than the measurement noise for the HET/HRS measurements (see Table 1). 
A periodogram analysis was carried out with a RV fitting code running a 
combination of a Lomb-Scargle (L--S) periodogram (Lomb 1976; Scargle 1982)
 and an iterative grid-search 
algorithm to locate the best fit to the data.  The interactive grid search 
is the method used to search parameter space for the best fitting model by 
adjusting model parameters until it converges on a solution.  Figure~5 
displays the L--S periodogram resulting from a Fourier analysis of the 
HD~102195 velocities. The horizontal dotted lines correspond to the 
false-alarm probabilities for a given L--S statistic (or power). A strong 
peak is found at a period of 4.11 days.  The false alarm probability 
associated with this peak is $<10^{-6}$.

We performed a Bayesian analysis of the RV data to identify if a planetary 
companion is present and determine their orbital parameters, following 
the methods in Ford (2005, 2006).  The RV variations are modeled as a 
single planet on a circular orbit. We assume a prior 
$p(P,K,\phi,\vec{C},\sigma_j) \sim \left[P (K+K_o) \sigma_j\right]^{-1}$, 
where $P$ is the orbital period, $K$ is the velocity semi-amplitude, 
$\phi$ is the orbital phase at the given epoch, $C_i$ is the $i^{\rm th}$ 
constant velocity offset, and $\sigma_j$ is the magnitude of the stellar 
jitter. We impose sharp cutoffs on the prior at $P_{\min} = 1$~day, 
$P_{\max} = 3$~year, $K_{\max} = 2$~km~s$^{-1}$, 
$\sigma_{j, \min} = 1$~m~s$^{-1}$, and $\sigma_{j,\max} =1$~km~s$^{-1}$.  
A modified Jeffery's prior is used for the velocity semi-amplitude to make 
the prior for $K$ normalizable, and we choose a scale parameter of 
$K_o=1$m s$^{-1}$. Our model also includes five velocity offsets, $C_i$, that 
are necessary due to the use of different velocity zero points for the 
various instruments and observing runs. We assume that each observation has 
an independent Gaussian observational uncertainty, $\sigma_{i,\mathrm{obs}}$, 
that is estimated from the photon statistics. We calculate the likelihood 
using an effective uncertainty of 
\hbox{$\sigma_{i,\mathrm{eff}} = \sqrt{\sigma_{i,\mathrm{obs}}^2+\sigma_j^2}$} 
for each observation.  To calculate the posterior probability distribution 
function, we analytically integrate over all but two of the model parameters 
using the Laplace and WKB approximations. Then, we directly integrate over 
the two remaining model parameters, $P$ and $\sigma_j$.

In Figure~6 we show the posterior probability density function (posterior PDF) 
marginalized over all model parameters except the orbital period, analyzing 
the observations from each observatory separately. The 14 observations with 
 ET at the KPNO 0.9~m coud\'e feed have a significantly larger measurement 
uncertainty than the other observations and do not isolate a single orbital 
solution (upper middle panel). Based on these observations, we obtained 
35 observations with the ET at the KPNO 2.1~m. Analysis of the velocities 
from the 2.1~m alone shows a $\ge99.9$\% posterior probability of having a periodic 
signal (compared with a constant velocity or quadratic trend), and a 
$\ge99.3$\% posterior probability integrated over all solutions contained in
the peak at 4.1 days (lower middle 
panel). Based on early observations with the 2.1~m, we obtained ten additional 
very high precision radial velocity measurements using the echelle 
spectrograph on HET. The HET observations by themselves (bottom panel) 
display the same periodicity as found in the KPNO/ET observations. Combining 
all the radial velocity observations (top panel), we find that there is a 
single dominant peak at 4.11~days that we interpret as the result of 
perturbations by a $m \sin i \simeq 0.49M_{J}$ planet.
 
In Figure~7, we show the marginalized posterior PDF combining all radial 
velocity observations. The posterior probability is dominated by one peak 
at $P = 4.11$~days that contains essentially all of the posterior probability. 
The peak near 1.3~days is a result of aliasing due to the day/night cycle 
and contains $\le10^{-9}$ of the posterior probability. The peak near 4.8~days 
shown in the coud\'e feed data (see inset) is the result of aliasing due to 
the lunar cycle (most observations occur near full moon) and contains only 
$\le10^{-18}$ of the posterior probability. Our best-fit solution has 
$P = 4.11453$~days, $K = 63.2$~m~s$^{-1}$, $\sigma_j = 5.9$~m~s$^{-1}$, and 
an rms of 16.0~m~s$^{-1}$.  We have also performed a Bayesian analysis 
using a full Keplerian model and the Markov chain Monte Carlo (MCMC) 
technique (Ford 2005). The present observations do not provide a significant 
detection of an orbital eccentricity, but do place upper limits on the 
allowed eccentricities. Marginalizing over all model parameters, we estimate 
that $\left<P\right>\simeq4.11434\pm0.00089$, 
$\left<K\right>\simeq63.4\pm2.0$m s$^{-1}$ and 
$\left<\sigma_j\right>\simeq5.8\pm1.8$~m~s$^{-1}$.  Our MCMC simulations 
find a 5\% posterior probability for all eccentricities larger than $0.096$, 
and a 0.1\% posterior probability for all eccentricities greater than 0.14. 
Table~5 summarizes the orbital parameters for HD~102195b.

An independent RV curve fitting was conducted using the {\it Gaussfit} 
software developed by part of the HST FGS science team (e.g., Jefferys et al. 
1987; McArthur et al. 1994; Cochran et al. 2004). This software uses a robust 
estimation method to find the combined orbital solution, where the offsets 
from different data sets are included as free parameters for fitting.  
Figure~8 shows the best-fit RV curve. The planet has a 4.1134$\pm$0.0009 day 
orbital period, which is consistent with the derived values from the L-S 
periodogram and also the Bayesian analysis. This analysis finds a solution 
with a minimum planet mass of 0.49~$M_{J}$ in a nearly circular orbit with 
an eccentricity of~0.06$\pm$0.02. These values are in excellent agreement 
with the Bayesian analysis above. The semimajor axis has a minimum size of 
0.049~AU.

The companion's mass is similar to other known hot-Jupiter exoplanets with
similar orbital periods around solar-type stars. The low eccentricity is not 
surprising since tidal circularization is expected for such a close planetary 
companion.

\subsection{Photometric Analysis}

Plots of the T10 APT photometric measurements from the 2004--05 and 2005--06
observing seasons are shown in Figures 9 and 10, respectively.  Although
the observations from 2004--05 are relatively few (38), the photometric
amplitude of HD~102195 was greater at that time.  The top panel of Figure~9
clearly shows cyclic variability with an amplitude of around 0.015 mag.
The power spectrum of these data is shown in the bottom panel of Figure~9
and gives a period of 12.3 $\pm$ 0.3 days, which we take to be the rotation
period of the star made apparent by rotational modulation in the visibility
of photospheric starspots.  The lightcurve closely resembles those of
other spotted stars (e.g., Henry, Fekel, \& Hall 1995).

The top panel in Figure 10 plots all the photometric data for the 2005--06 
observing season. A power spectrum of these second season observations 
exhibits the same 12.3-day periodicity but with much lower amplitude.  The 
observations in this figure are plotted against planetary orbital phase 
computed from the orbital elements in Table 5; zero phase refers to a time 
of inferior conjunction (mid transit).  A least-squares sine fit to the 
brightness measurements in the top panel gives a semi-amplitude of only 
0.0004$\pm$0.0002 mag.  Given the cycle-to-cycle variation in the lightcurve 
caused by a continually evolving starspot distribution, this small amplitude 
on the radial velocity period is consistent with {\it no} brightness 
variability on that period.  Thus, the photometry supports planetary reflex 
motion as the cause of the radial velocity variability. 

Since the 2005--06 brightness measurements are both more numerous and 
exhibit lower variability than the observations of the previous season, 
the observations from the second season are more suitable for seeking 
possible transits of the planet across the disk of the star.  The solid 
curve in each panel of Figure~10 approximates the predicted transit 
lightcurve, assuming a planetary orbital inclination of 90\arcdeg (central 
transits).  The out-of-transit light level corresponds to the mean brightness 
of the observations.  The transit duration is estimated from the orbital 
elements, while the transit depth is derived from the estimated stellar 
radius (Table~3) and an assumed planetary radius equal to Jupiter's.  The 
bottom panel of Figure~10 shows the observations around the predicted time 
of mid transit replotted with an expanded scale on the abscissa.  The 
horizontal error bar below the predicted transit curve represents the 
estimated uncertainty in the time of mid transit, based on the orbital 
elements.  The vertical error bar represents the 0.0019 mag precision of the 
observations.

These 2005--06 photometric observations of HD~102195 cover the predicted 
transit window quite well and reveal no sign of a transit.  The 47 
observations within the transit window in the bottom panel of Figure~10 
have a mean brightness level of 0.3683 $\pm$ 0.0005 mag.  The 383 
observations outside the transit window have a mean of 0.3676 $\pm$ 0.0002 
mag.  Therefore, the mean brightness values inside and outside of the transit 
window agree to within 0.0007 mag, i.e., to within their respective 
uncertainties.  Our photometric observations rule out even very shallow 
transit events.  

\section{Discussion}

HD 102195 is a mildly active G8V dwarf. Stellar activity, such as starspots, 
introduces jitter into the RV measurements. The amplitude of the stellar 
jitter caused by stellar activity can be estimated from the previously 
derived formula of Santos et al. (2000). Using the measured chromospheric 
activity index, $R'_{HK} = 5\times 10^{-5}$, we estimate that the RMS RV 
jitter should be $\sim$ 19~m~s$^{-1}$.  This level is consistent with 
the estimate from the Bayesian analysis of the radial velocity observations. 
Based on previous RV studies of active G dwarfs with photometric variations 
similar to HD 102195 (e.g., Paulson et al. 2004; Santos et al. 2000), we 
expect that the semiamplitude of the starspot induced radial velocity 
variation for HD~102195 may be on the order of $\sim$ 10--20 m s$^{-1}$. 
Even if the photometric and RV periods were equal, the deduced level of 
stellar activity could not account for the observed RV variations in 
HD~102195.  This strengthens the conclusion from \S 5.5 above that stellar 
activity is not responsible for the 4.11-day radial velocity variation.

It is unusual (and of some concern) that the 12.3-day photometric period 
appears to be exactly (at least within the measurement errors) three times 
that of the 4.11-day radial velocity period. Could three similar starspot 
groups spaced approximately 120\arcdeg ~in longitude on the star be 
responsible for the observed radial velocity period?  Figure~9 shows
that there is essentially no power in the brightness variations at
4.11 days (corresponding to a frequencey of 0.24 cycles per day).  
Conversely, an examination of Figure~7 shows that there is essentially no 
power in the radial velocity variations at the photometric period of 
12.3~days.  This mismatch of the photometric and radial velocity periods
indicates that they arise from separate causes and supports the
planetary interpretation. 

The Ca II H and K emisison measurements also support the planetary
interpretation of the RV variations in HD 102195. The standard technique
for analyzing stellar H and K emission is the S index described by Duncan
et al. (1991) This quantitiy is defined by Duncan et al. as S=(H+K)/(V+R)
where V and R are the summed flux in un-normalized 20 \AA\ wide continuum
bands extending from 3891-3911 \AA\ and 3991-4011 \AA\ respectively and H and K are
channels centered on the cores of the Ca II H and K and have a measured
triangular instrument response of FWHM 1.09 \AA. For our S index
measurements using spectra obtained at the KPNO 0.9m coude' telescope we
define the quantities V and R in the same way as Duncan et al. (1991) ,
but differ in our definition of the quantities H, K which we define as the
integrated flux in 1.4 \AA\ bands centered of the Ca II H and K lines. We
call the resulting S index S$_{coude}$ to differentiate it from the Mt Wilson
S index. No attempt has been made to put S$_{coude}$ on the Mt. Wilson scale,
since we are only interested in the variability of S$_{coude}$. The analysis
of the chromospheric activity traced by the S index does not show any
variability at the 4.11 day period of the planet, but does display longer
term changes that appear consistent with the photometric rotation period.
The mean S$_{coude}$ index for the entire run is measured to be 0.024, and
varies from 0.023 to 0.025. 
For comparison, in the reference star, Tau
Ceti, the S index measuremtn varies less than 0.0005. This result further
suggests that stellar activity is not the source of the detected RV
signals.

In addition, the stellar rotation period measured from the photometry is 
consistent with the projected rotation velocity measured from the spectral 
analysis of the SARG and FOCES high resolution spectra. The measured 
\hbox{$v \sin i = 3.23 \pm 0.07$ km~s$^{-1}$} corresponds to a maximum rotation 
period of $13.1 \pm 0.3$~days if a stellar radius of 0.84~R$_\odot$ is 
adopted. This does indicate that the stellar rotation axis may be close to
an inclination of 90\arcdeg. However, as discussed in \S 5.5 above, 
photometric transits have been ruled out by our photometry.

This planetary detection is the first time an extrasolar planet has been
discovered via RV variations around a star fainter than $V = 8$ with a 
sub-meter aperture telescope.  This discovery was made possible by the high
throughput of the ET instrument. The total measured system detection
efficiency of~18\% (from the telescope to the detector) or 49\% (from the 
fiber output to the detector) is about four times higher than has been 
achieved with HARPS, a state-of-the-art echelle spectrograph being used 
on the ESO 3.6-m telescope (Pepe et al. 2002). 

Future DFDI instruments promise additional capabilities.  Due to the use of 
the single-order medium resolution stellar spectra superimposed with 
interference fringes, this method can be easily modified to allow 
multiple-object Doppler measurements (Ge 2002). A prototype multi-object 
DFDI obtained data in the spring of 2005 with the SDSS telescope and 
demonstrated this feasibility (Ge et al. 2005). 

In addition, for single-object observations, the single-order medium 
resolution spectrograph can be replaced with a cross-dispersed echellette 
spectrograph to cover multiple orders of dispersed fringes to gain wavelength 
coverage. The echellette can also increase the spectral resolution due to 
its higher dispersion over the low single-order blazed grating. Since the 
Doppler sensitivity is roughly proportional to the square root of the 
spectral resolution and wavelength coverage (Ge~2002), it is likely that an 
ET interferometer coupled with a cross-dispersed echellette will increase 
the Doppler sensitivity by a factor of $\sim$3--5 times over current designs.

In summary, the detection of a planet around HD 102195 with a DFDI 
instrument demonstrates that the new-generation, high-throughput 
interferometric Doppler method is an effective technique for identifying 
extrasolar planets.  This approach offers a number of opportunities for 
high precision RV measurements and planet surveys, especially for 
stars fainter than previous surveys with echelle Doppler instruments have
reached.  This technique may become a significant component in the effort 
to assemble the large sample of extrasolar planets required for a 
comprehensive characterization of these objects.

\acknowledgments

We are grateful to Stuart Shaklan, Michael Shao, Chas Beichman, Richard 
Green, Skip Andree, Daryl Wilmarth, and the KPNO staff for their generous 
support and advice and to Larry Molnar for his help in the early 
photometric data acquisition. The HET observations, whose timings were 
critical, were made possible by the dedicated efforts of the HET staff.  We 
are indebted to Sara Seager and Eric Agol for a number of stimulating 
discussions. We thank Aldo Martinez Fiorenzano for helping the line bisector analysis of
HD 102195. We appreciate many valuable comments made by the referee, which helped to 
improve the paper quality. This work is supported by the National Science Foundation 
grant AST 02-43090, JPL, the Pennsylvania State University and the University 
of Florida.  SM and JvE acknowledge the JPL Michelson Fellowship funded 
by NASA. GWH is greatful for the effors of Lou Boyd at Fairborn Observatory
and acknowledges support from NASA grant NCC5-511 and NSF grant 
HRD 97-06268.  E.B.F. acknowledges the support of the Miller Institute for 
Basic Research. WDC, ME, and RAW were supported by NASA grants NNG04G141G 
and NNG05G107G to The University of Texas at Austin.  The Hobby-Eberly 
Telescope (HET) is a joint project of the University of Texas at Austin, 
the Pennsylvania State University, Stanford University, 
Ludwig-Maximillians-Universit\"at M\"unchen, and 
Georg-August-Universit\"at G\"ottingen. The HET is named in honor of its 
principal benefactors, William P. Hobby and Robert E. Eberly. DM was supported by the Universidad Complutense de Madrid and the Spanish
Ministerio de Educaci\'on y Ciencia (MEC),  Programa Nacional de
Astronom\'{\i}a y Astrof\'{\i}sica under grant AYA2005-02750. DM is grateful to Raquel M. Mart\'{\i}nez, Jes\'us Maldonado and Benjam\'{\i}n
Montesinos for their help during the observations. FOCES high resolution optical spectra were 
based on observations collected at the Centro Astron\'omico Hispano
Alem\'an (CAHA) at Calar Alto, operated jointly by the Max-Planck Institut
f$\ddot{\rm u}$r Astronomie and the Instituto de Astrof\'{\i}sica de
Andaluc\'{\i}a (CSIC).
This research 
has made the use of the SIMBAAD data base, operated at ADC, Strasbourg, France.

\clearpage
\begin{figure}
\epsscale{.80}
\plotone{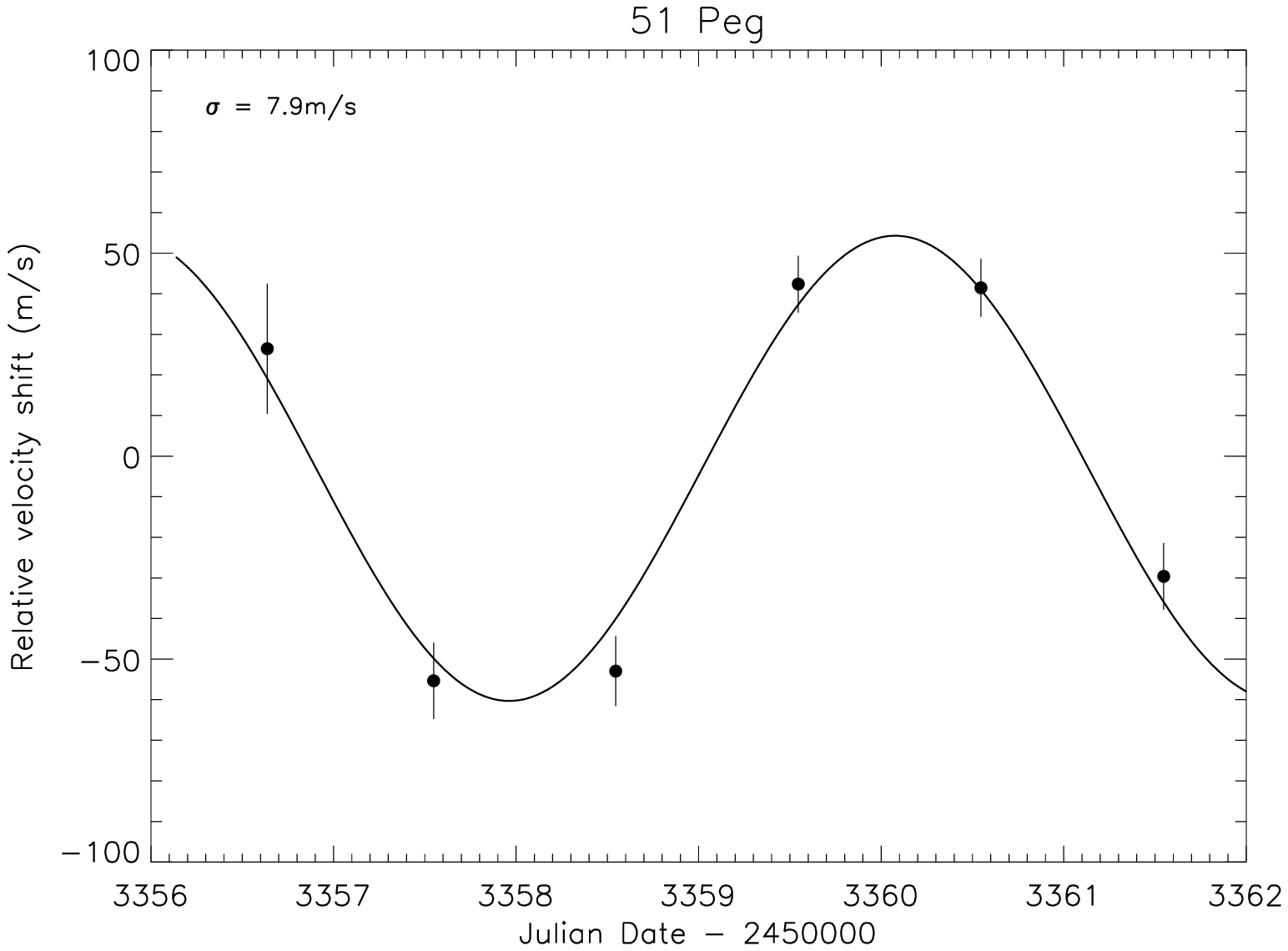}
\caption{RV measurements of 51~Peg taken with ET at the 0.9~m coud\'e feed 
in 2005 May.  The solid line is the predicted RV curve based on the orbital solution in  
Marcy et al. (1997). 
The exposure times were 15 minutes and the typical RV error for each point 
is 7.9~m~s$^{-1}$.}
\label{fig1}
\end{figure}

\clearpage
\begin{figure}
\epsscale{.80}
\plotone{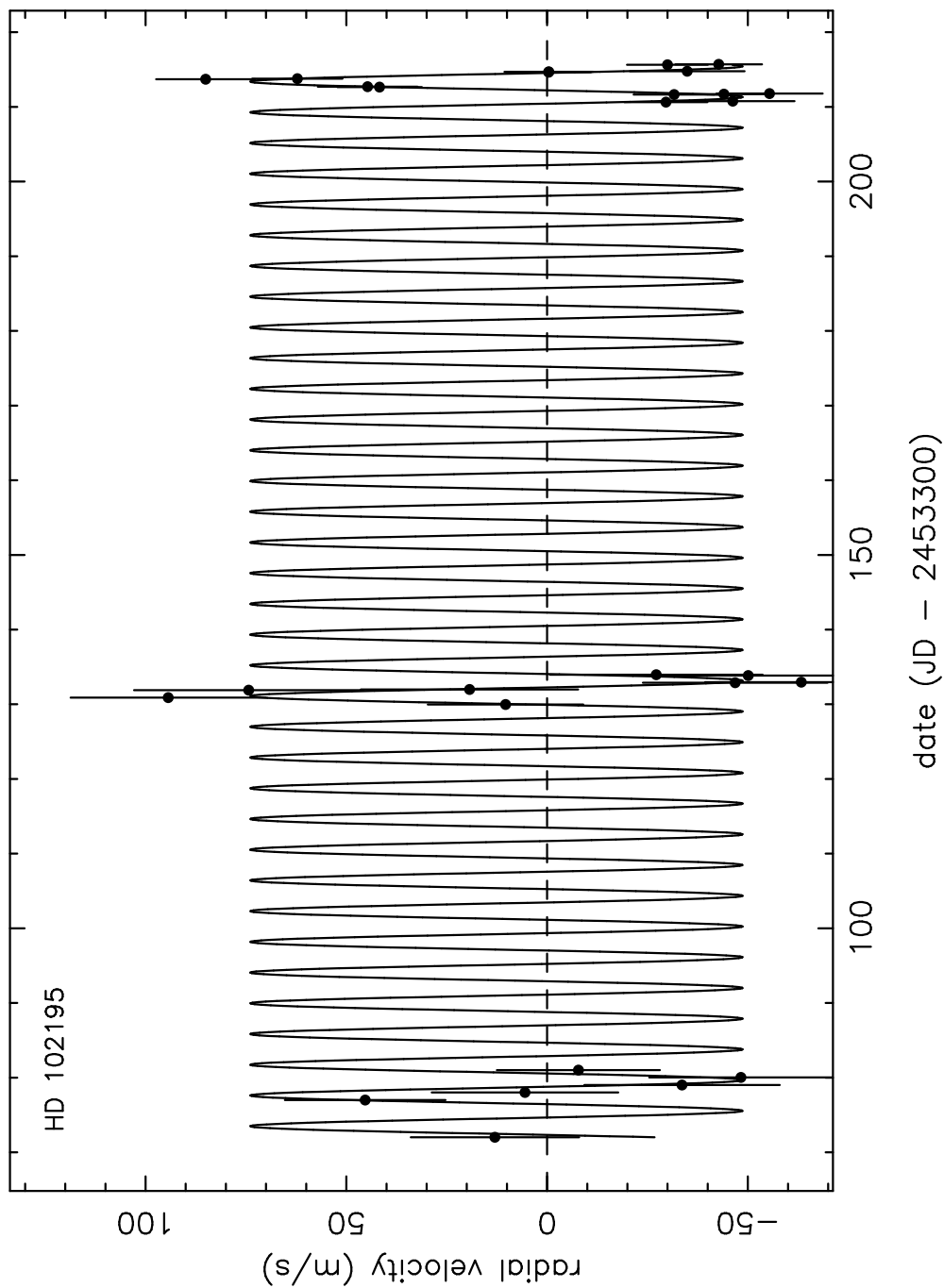}
\caption{Radial velocities for HD~102195, showing the 28 measurements from 
ET at the KPNO coud\'e feed/2.1~m in the spring of 2005.  The rms errors 
for the ET measurements at KPNO are $\sim$10 m s$^{-1}$ and 
$\sim$20 m s$^{-1}$ for the 2.1~m and coud\'e feed, respectively.  The 
solid line is the Keplerian orbital fit to the observed data.} 
\label{fig2}
\end{figure}

\clearpage
\begin{figure}
\epsscale{.80}
\plotone{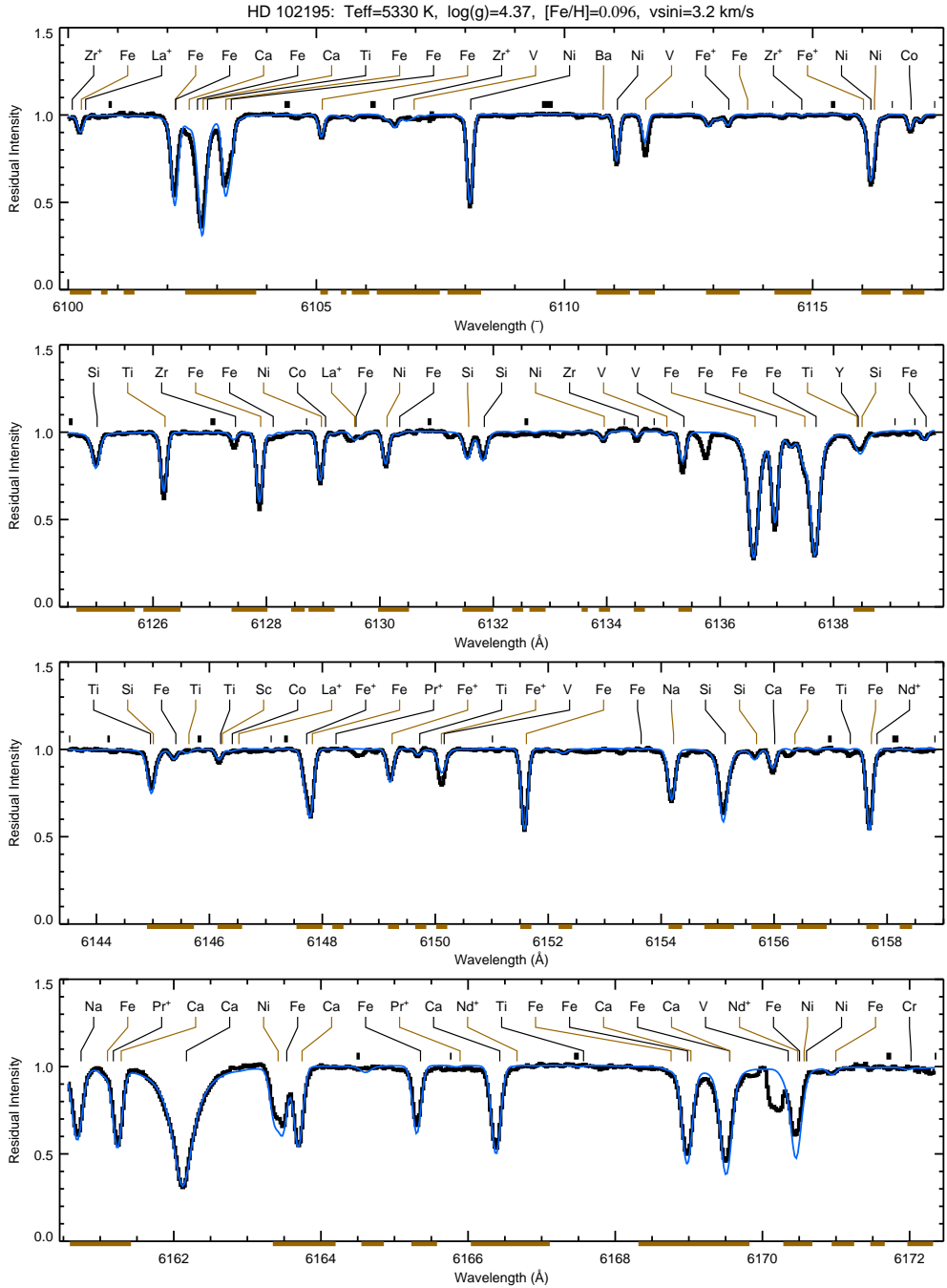}
\caption{Part of the combined normalized observed spectrum and superimposed 
synthetic spectrum of HD~102195 for four different wavelength intervals. 
The black color lines are the observed data. The grey lines are the synthetic. 
The spectral resolution is $R=150,000$ ($\sim$0.02 \AA\ per pixel) and an 
average S/N $\sim$ 300 per pixel in the combined continuum spectrum.  The 
synthetic spectrum is produced by a model using the stellar parameters in 
Table 3.}
\label{fig3}
\end{figure}

\clearpage
\begin{figure}
\epsscale{.80}
\plotone{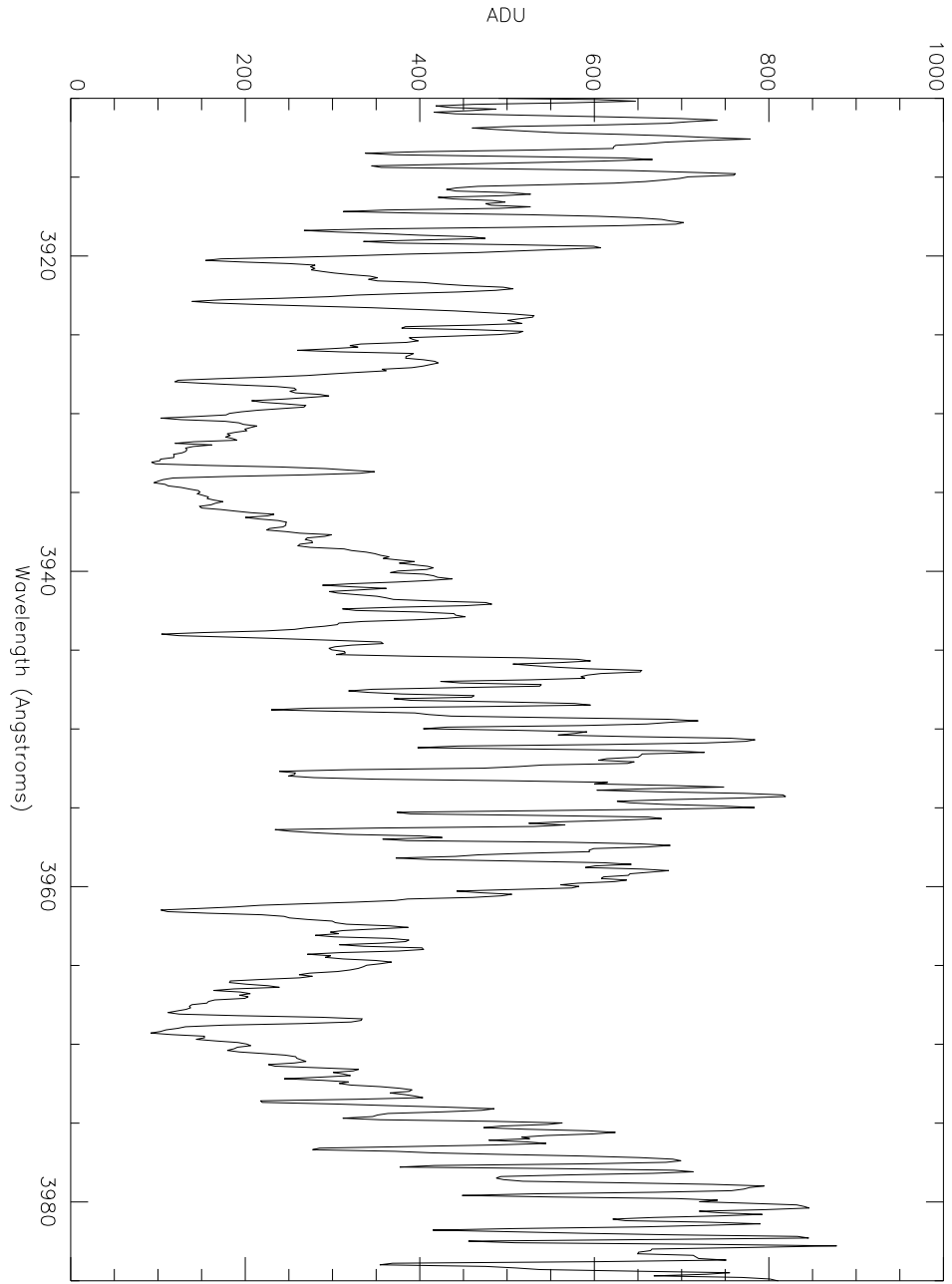}
\caption{A typical Ca II H and K emission line spectrum of HD 102195 taken with the KPNO 
0.9~m coud\'e long-slit spectrograph after the background is subtracted and 
the flux is summed.  The spectral resolution is $R=10,000$.  H and K emission 
lines are clearly visible in the core of the broad H and K absorption lines.}
\label{fig4}
\end{figure}
 
\clearpage
\begin{figure}
\epsscale{.80}
\plotone{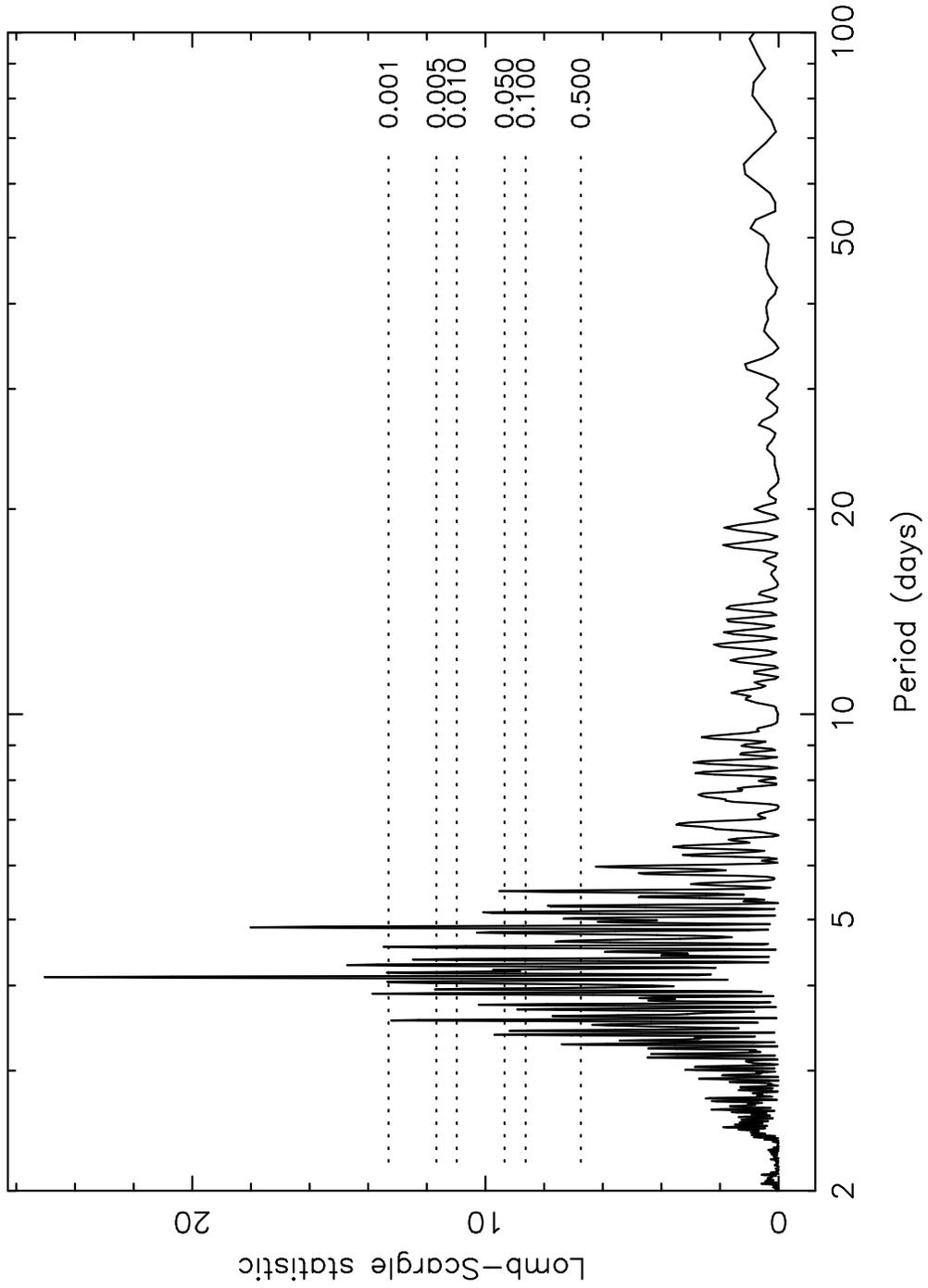}
\caption{Periodogram of the velocities for HD~102195, showing peak power at 
4.11 days with a false alarm probability $\sim10^{-6}$.}
\label{fig5}
\end{figure}

\clearpage
\begin{figure}
\epsscale{.80}
\plotone{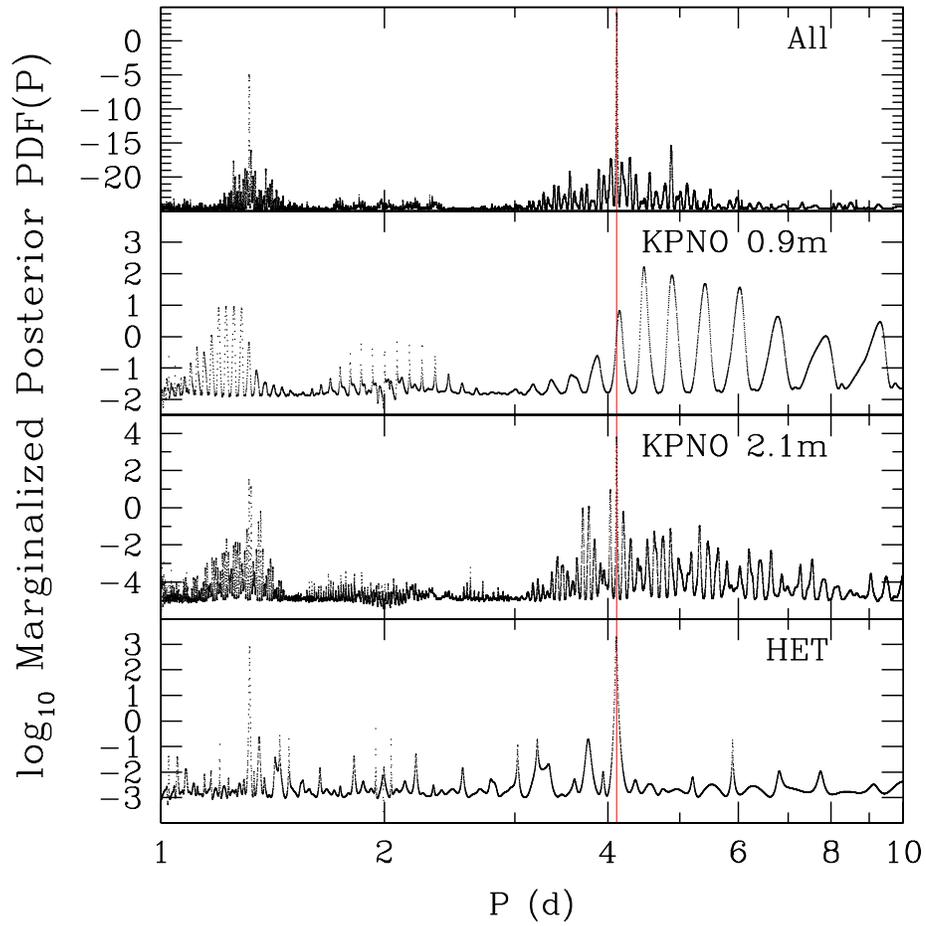}
\caption{The posterior probability density function marginalized over all 
model parameters except for the orbital period for RV data taken at different 
telescopes.}
\label{fig6}
\end{figure}

\clearpage
\begin{figure}
\epsscale{.80}
\plotone{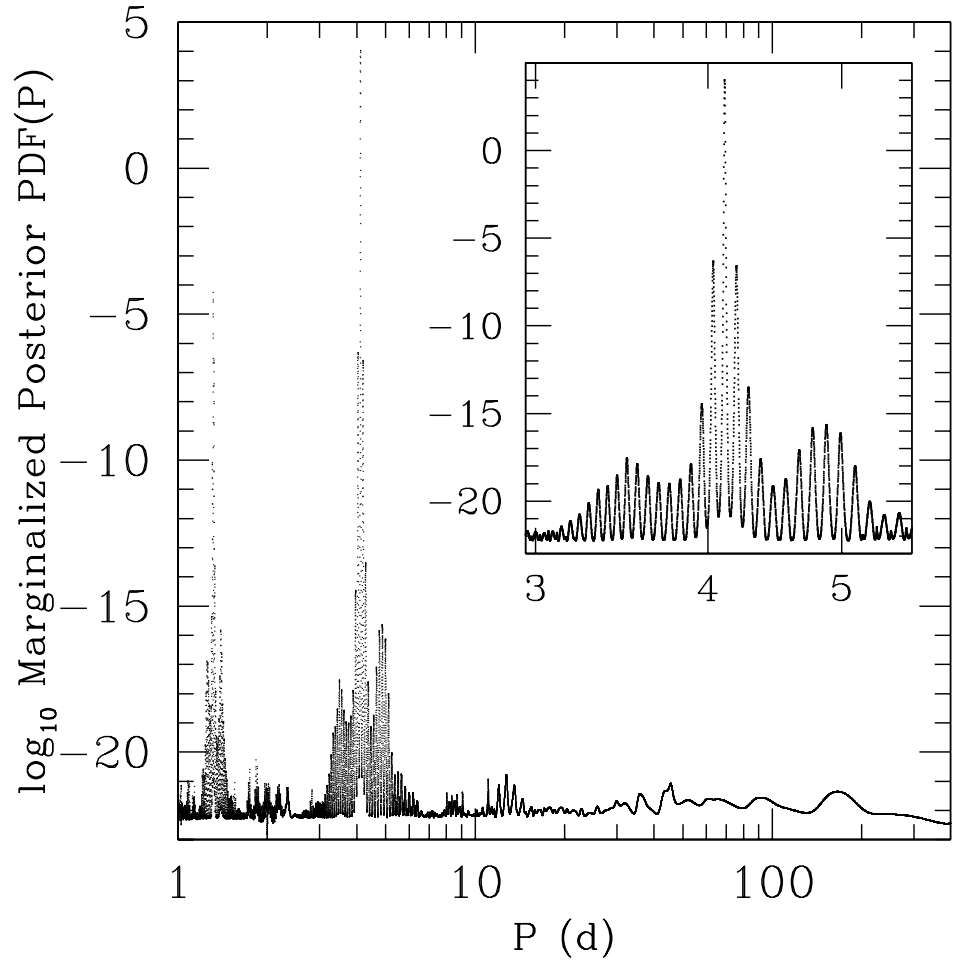}
\caption{The marginalized posterior PDF combining all RV data taken at the 
coud\'e feed, 2.1~m and the HET. Nearly all of the power is in the 4.11-day 
period; there is no indication of any significant RV variation at the 
12.3-day photometric period.}
\label{fig7}
\end{figure}

\clearpage
\begin{figure}
\epsscale{.80}
\plotone{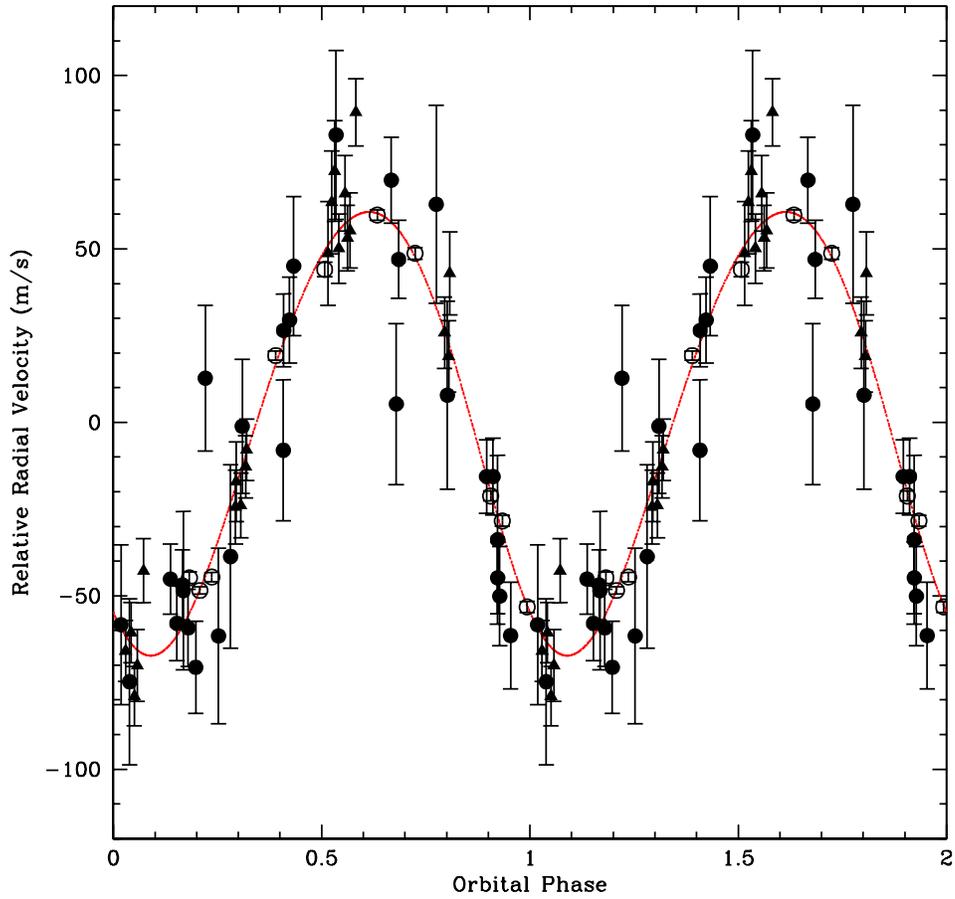}
\caption{Phased radial velocities for HD~102195; the KPNO coud\'e data are 
filled circles, KPNO 2.1~m data are filled triangles, and the HET data are 
open circles. Two orbital cycles are shown and each observation is plotted 
as two points.  The best fit to the data yields an orbital period of 
4.11434~days and a velocity semiamplitude of 63.4~m~s$^{-1}$. If the stellar 
mass is 0.93M$_\odot$, the derived minimum planetary mass is 0.488~$M_J$ 
and the minimum orbital radius is 0.0491~AU.}
\label{fig8}
\end{figure}

\begin{figure}[t!]
\figurenum{9}
\epsscale{0.85}
\plotone{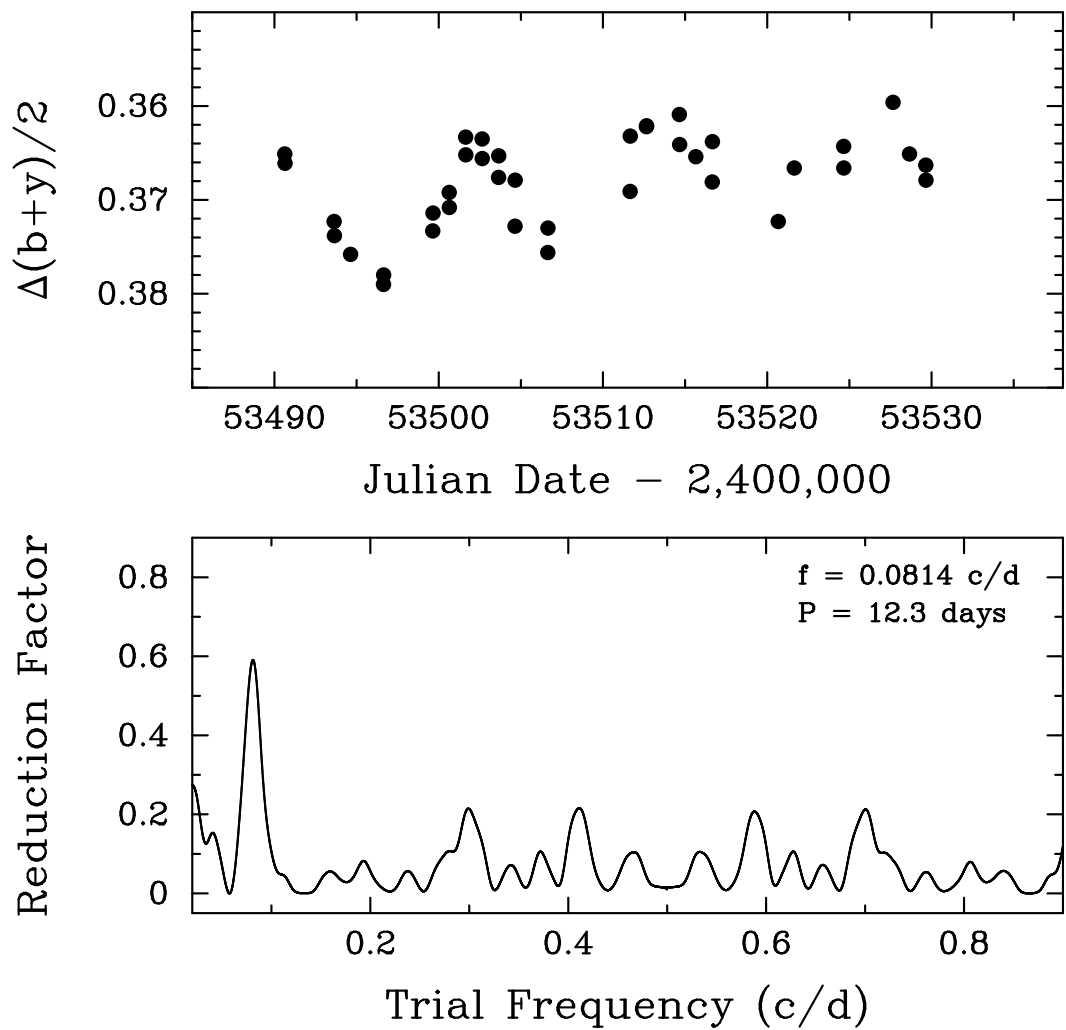}
\figcaption{{\it Top}:  The Str\"omgren $(b+y)/2$ photometric observations of 
HD~102195 obtained at the end of the 2004--05 observing season with the T10 
0.8~m APT at Fairborn Observatory.  A cyclic variation of about 0.015 mag 
in the brightness of the star is easily seen.  {\it Bottom}:  The power
spectrum of the observations in the top panel.  A clear period of 12.3 days
is interpreted as the rotation period of the star.}
\label{fig9}
\end{figure}

\begin{figure}[t!]
\figurenum{10}
\epsscale{0.85}
\plotone{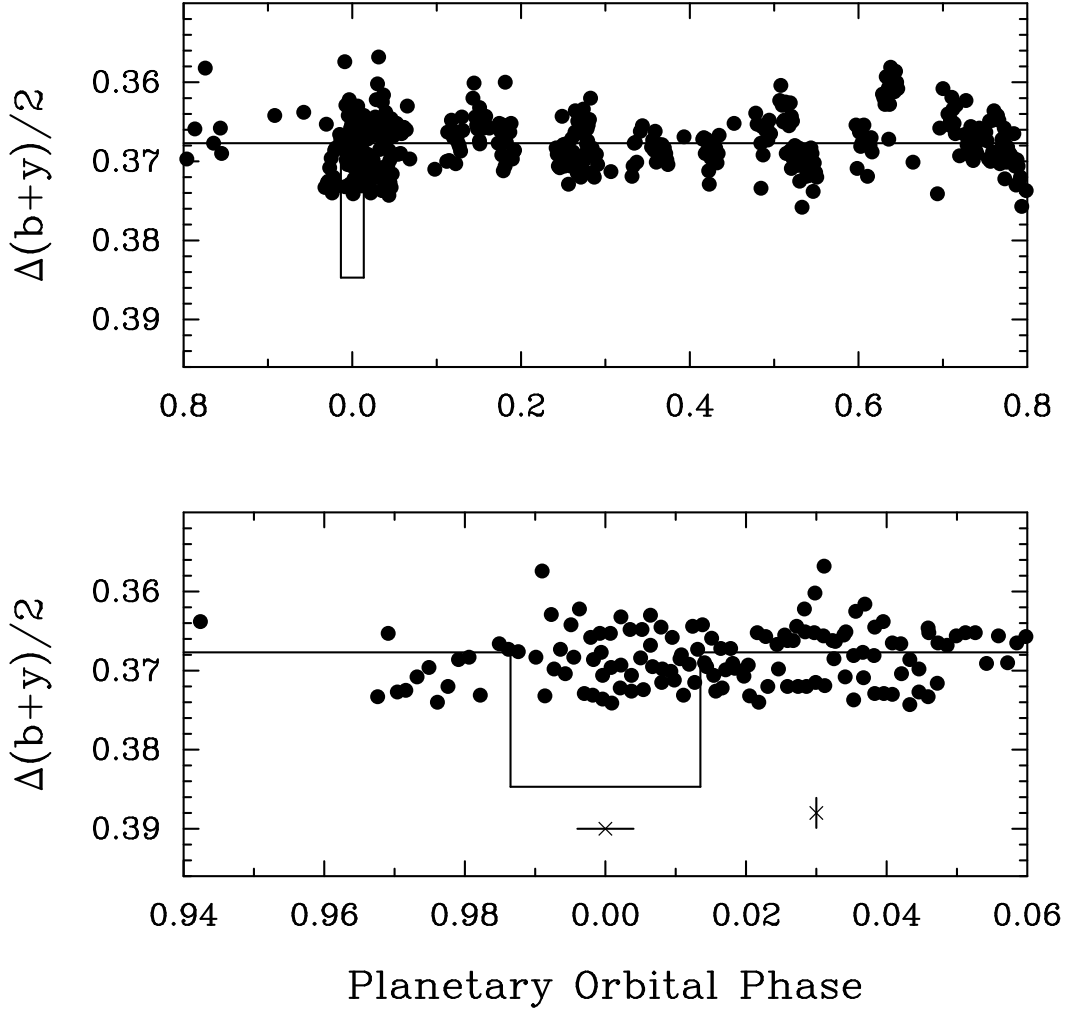}
\figcaption{{\it Top}:  The Str\"omgren $(b+y)/2$ photometric observations 
of HD~102195 obtained during the 2005--06 observing season with the 
T10 0.8~m APT at Fairborn Observatory and plotted against orbital phase of 
the planetary companion.  The predicted time, depth, and duration of possible 
transits are shown schematically.  The star exhibits no optical variability 
on the radial velocity period larger than 0.0004 mag or so.  {\it Bottom}:  
The observations around the predicted time of transit are replotted with
an expanded scale on the abscissa.  The error bars are described in the
text.  Even very shallow transits are ruled out by these observations.}
\label{fig10}
\end{figure}

\begin{figure}[t!]
\figurenum{11}
\epsscale{0.85}
\plotone{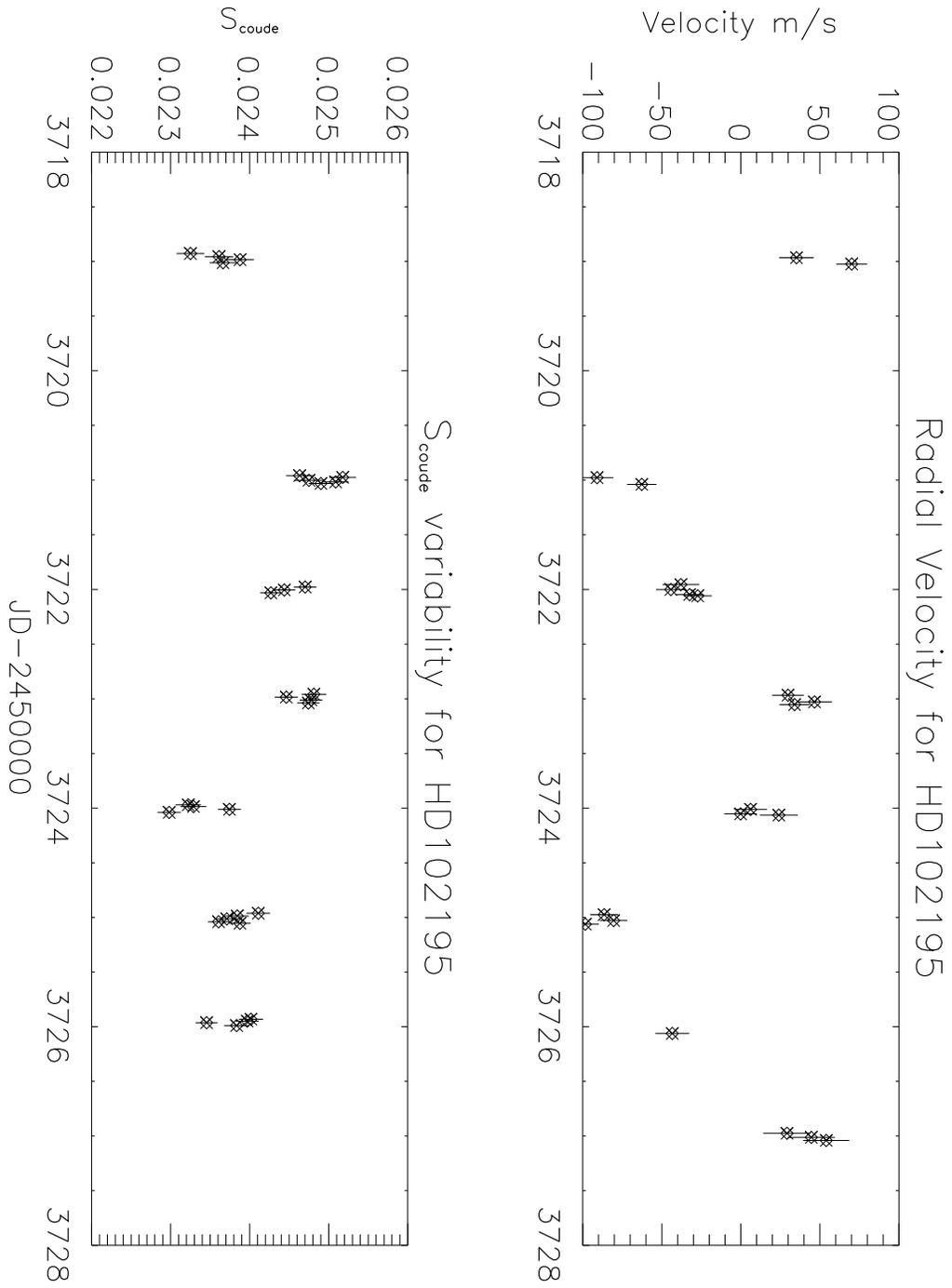}
\figcaption{The measurements of the 
S$_{coude}$ index ({\it bottom}) and RV ({\it top}) during the same observation run at Kitt Peak.}
\label{fig11}
\end{figure}









\clearpage
\begin{flushleft}
Table 1. Radial Velocities for HD 102195
\end{flushleft}
\begin{tabular}{llrrllrr} \hline\hline
\ \ \ \ \ JD &Instrument$^a$ &RV &errors &
\ \ \ \ \ JD &Instrument$^a$ &RV &errors \\
$-2450000$ && m s$^{-1}$ & m s$^{-1}$ &
$-2450000$ && m s$^{-1}$ & m s$^{-1}$\\
\hline
&&&\\
 3372.014& CF, BK2 & 133.0 & 21.0 &
3696.034& HET & $-$60.6& 1.3\\
 3376.999& CF, BK2 & 165.3& 20.0&
3697.034& HET & $-$52.0& 1.1\\
 3378.011& CF, BK2 & 125.5& 23.2&
3701.029& HET & $-$55.9&0.8\\
 3379.013& CF, BK2 & 86.4& 24.3&
3704.016& HET & $-$35.8& 1.3\\
 3380.024& CF, BK2 & 71.7& 22.8&
 3718.966& 2.1m & 56.0& 10.9\\
 3381.011& CF, BK2 & 112.2& 20.3&
 3719.023& 2.1m & 90.0& 9.7\\
 3429.966& CF, BK4 & $-$99.6& 19.3&
 3720.978& 2.1m& $-$69.5& 10.3\\
 3430.892& CF, BK4 & $-$15.6& 24.3&
 3721.039& 2.1m & $-$42.1& 9.2\\
 3431.882& CF, BK4 & $-$35.7& 28.5&
 3721.956& 2.1m & $-$16.5& 11.5\\
 3431.991& CF, BK4 & $-$90.7& 27.1&
 3722.001& 2.1m & $-$23.3& 9.3\\
 3432.881& CF, BK4 & $-$156.9& 23.0&
 3722.048& 2.1m & $-$12.2& 9.0\\
 3432.966& CF, BK4 & $-$173.3& 24.0&
 3722.059& 2.1m & $-$7.2& 8.9\\
 3433.844& CF, BK4 & $-$160.1& 25.3&
 3722.967& 2.1m & 50.7& 10.0\\
 3433.963& CF, BK4 & $-$137.2& 26.5&
 3723.029& 2.1m & 66.6& 10.9\\
 3510.641& 2.1m, BK5 & $-$29.6& 10.3&
 3723.053& 2.1m & 53.7& 9.5\\
 3510.769& 2.1m, BK5 & $-$46.2& 15.4&
 3724.012& 2.1m & 26.5& 10.3\\
 3511.643& 2.1m, BK5 & $-$31.6& 10.1&
 3724.053& 2.1m & 19.6& 10.3\\
 3511.699& 2.1m, BK5 & $-$44.0& 11.1&
 3724.064& 2.1m & 43.6& 11.9\\
 3511.774& 2.1m, BK5 & $-$55.4& 13.2&
 3724.975& 2.1m & $-$65.3& 8.8\\
 3512.641& 2.1m, BK5 & 41.7& 10.5&
 3725.029& 2.1m & $-$60.0& 8.7\\
 3512.699& 2.1m, BK5 & 44.7& 12.4&
 3725.062& 2.1m & $-$78.3& 8.6\\
 3513.704& 2.1m, BK5 & 85.0& 12.3&
 3726.062& 2.1m & $-$23.8& 10.6\\
 3513.777& 2.1m, BK5 & 62.2& 11.2&
 3726.975& 2.1m & 49.3& 14.9\\
 3514.646& 2.1m, BK5 & $-$0.4& 10.6&
 3727.012& 2.1m & 64.1& 14.8\\
 3514.708& 2.1m, BK5 & $-$0.4& 11.0&
 3727.040& 2.1m & 73.1& 14.5\\
 3514.775& 2.1m, BK5 & $-$34.9& 14.3&
3731.949& HET & 41.3& 1.4\\
 3515.638& 2.1m, BK5 & $-$30.0& 10.1&
3737.946& HET & $-$52.2& 1.3\\
 3515.700& 2.1m, BK5 & $-$42.7& 10.8&
3740.922& HET & $-$28.6& 1.1\\
3694.035& HET & 36.7& 1.8&
3742.912& HET & 11.9& 1.1\\
& & & &
3743.915& HET & 52.3& 1.1\\

&&&\\
\hline
\end{tabular}

$^a$ CF = 0.9-m Coude Feed; BK = Block

\clearpage
\begin{deluxetable}{ccc}
\tablenum{2}
\tablewidth{0pt}
\tablecaption{PHOTOMETRIC OBSERVATIONS OF HD~102195}
\tablehead{
\colhead{Observation Date} & \colhead{$\Delta (b+y)/2$} \\
\colhead{(HJD $-$ 2,400,000)} & \colhead{(mag)}
}
\startdata
53,490.6339 & 0.3651 \\
53,490.6494 & 0.3661 \\
53,493.6380 & 0.3723 \\
53,493.6498 & 0.3738 \\
53,494.6381 & 0.3758 \\
\enddata
\tablecomments{Table 2 is presented in its entirety in the electronic edition
of the Astrophysical Journal.  A portion is shown here for guidance regarding
its form and content.}
\end{deluxetable}

\clearpage
\begin{flushleft}
Table 3. Stellar Parameters for HD 102195
\end{flushleft}
\begin{tabular}{ll} \hline\hline
Parameter & Value \\
\hline
&\\
$V$& 8.05\\
$M_V$& 5.73\\
$B-V$ & 0.84\\
Spectral type & G8V\\
Distance & 29 pc \\
$[$Fe/H$]$ &$ 0.096 \pm 0.032$ \\
$T_{\rm eff}$ &$5330 \pm 28$ K\\
$v \sin i$ &3.23$\pm0.07$ km s$^{-1}$\\
$\log g$ & 4.368$\pm$0.038 [log(cm s$^{-2}$)]\\
BC & $-$0.177\\
$L_{\rm star}$ &  0.463$\pm$0.034 L$_\odot$\\
$M_{\rm star}$ & 0.926$\pm$0.016 $M_\odot$\\
$R_{\rm star}$ &0.835$\pm$0.016 $R_\odot$\\
log $R'_{HK}$&$-4.30$ \\
$P_{\rm rot}$ & 12.3$\pm0.2$ days\\
Age & 0.6-4.2 Gyr\\
&\\
\hline

\end{tabular}

\clearpage

\begin{flushleft}
Table 4. Equivalent width (EW) and surface flux (logF$_{\rm S}$) of the
different chromospheric activity indicators 
\end{flushleft}
\begin{flushleft}
\scriptsize
\begin{tabular}{lcccccccccccccccc}
\noalign{\smallskip} \hline \hline
\noalign{\smallskip}
  &     \multicolumn{6}{c}{$EW$(\AA) in the subtracted spectrum} & &
         \multicolumn{6}{c}{logF$_{\rm S}$ (erg cm$^{-2}$ s$^{-1}$)} \\
\cline{2-7} \cline{9-14}
\noalign{\smallskip}
Date & \multicolumn{2}{c}{Ca~{\sc ii}} & & \multicolumn{3}{c}{Ca~{\sc
ii}IRT} & & \multicolumn{2}{c}{Ca~{\sc ii}} & & \multicolumn{3}{c}{Ca~{\sc
ii} IRT}
\\
\cline{2-3} \cline{5-7} \cline{9-10} \cline{12-14}
\noalign{\smallskip}
 & K   & H & H$\alpha$ &
$\lambda$8498 & $\lambda$8542 & $\lambda$8662 & & K   & H & H$\alpha$ &
$\lambda$8498 & $\lambda$8542 & $\lambda$8662
\\

\noalign{\smallskip}
\noalign{\smallskip}
\hline
\noalign{\smallskip}
01/14/06   & 0.313 & 0.164 & 0.045 & 0.084 & 0.126 & 0.122 &  & 5.75 & 6.04 & 5.
29 & 5.43 & 5.61 & 5.59 \\
%
\noalign{\smallskip}
\hline

\end{tabular}
\end{flushleft}

\clearpage

\begin{flushleft}
Table 5. Orbital Parameters for HD 102195b
\end{flushleft}
\begin{tabular}{ll} \hline\hline
Parameter & Value \\
\hline
&\\
$P$ & 4.11434$\pm$0.00089 days \\
$T_p$ & 2453732.7$\pm$0.5\\
$e$ & $<$0.14 \\
$a$ & 0.0491 AU \\
$\omega$ (deg)& 143.4$\pm$15.4 \\
$K$ & 63.4$\pm$2.0 m s$^{-1}$\\
$m \sin i $ & 0.488 $\pm$ 0.015 $M_J\, \frac{M_*}{0.93M_{\odot}}$ \\
$\sigma_j$ & 5.8$\pm$1.8 m~s$^{-1}$ \\ 
rms & 16.0 m~s$^{-1}$ \\
\hline

\end{tabular}

\end{document}